\documentclass[aps,prd,11pt,preprint,nofootinbib,showkeys,preprintnumbers]{revtex4-1}
\usepackage{amsmath,amssymb,amsthm,mathrsfs,bbm}
\usepackage{latexsym,amscd,amsbsy,amsfonts,dsfont}
\usepackage{graphicx}
\usepackage[utf8]{inputenc}
\usepackage{subfigure}  
\usepackage{float}
\usepackage{slashed}
\usepackage{cancel}
\usepackage{multirow}
\usepackage{soul}
\usepackage{bm}
\usepackage{physics}
\usepackage[normalem]{ulem}
\usepackage{comment}
\usepackage[usenames,dvipsnames]{xcolor}
\usepackage[
      colorlinks=true,
      linktocpage=true,
      breaklinks=true,
      linkcolor=Aquamarine,
      urlcolor=Purple,
      filecolor=black,
      citecolor=Purple,
      menucolor=black,
      pdfstartview=FitV,
      bookmarksopen=true
      ]{hyperref}

\makeindex

\begin{document}
\count\footins = 1000

\title{No-hair and almost-no-hair results for static axisymmetric black holes and ultracompact objects in astrophysical environments}

\author{Carlos Barceló}
\email{carlos@iaa.es}
\affiliation{Instituto de Astrofísica de Andalucía (IAA-CSIC), Glorieta de la Astronomía, 18008 Granada, Spain}
\author{Ra\'ul Carballo-Rubio}
\email{raul@sdu.dk}
\affiliation{CP3-Origins, University of Southern Denmark, Campusvej 55, DK-5230 Odense M, Denmark}
\author{Luis J. Garay}
\email{luisj.garay@ucm.es}
\affiliation{Departamento de F\'{\i}sica Te\'orica and IPARCOS, Universidad Complutense de Madrid, 28040 Madrid, Spain}
\author{Gerardo Garc\'ia-Moreno}
\email{ggarcia@iaa.es}
\affiliation{Instituto de Astrof\'{\i}sica de Andaluc\'{\i}a (IAA-CSIC), Glorieta de la Astronom\'{\i}a, 18008 Granada, Spain}

\preprint{IPARCOS-UCM-23-115}

\begin{abstract}
    No-hair theorems are uniqueness results constraining the form of the metric of black holes in general relativity. These theorems are typically formulated under idealized assumptions, involving a mixture of local (regularity of the horizon) and global aspects (everywhere vacuum spacetime and asymptotic flatness). This limits their applicability to astrophysical scenarios of interest such as binary black holes and accreting systems, as well as their extension to horizonless objects. A previous result due to G\"urlebeck constrains the asymptotic multipolar structure of static spacetimes containing black holes surrounded by matter although not revealing the possible structure of the metric itself. In this work, we disentangle some of these assumptions in the static and axisymmetric case. Specifically: i) we show that only a one-parameter family of black-hole geometries is compatible with a given external gravitational field, ii) we also analyze the case in which the central object is close to forming an event horizon but is still horizonless and show that the deviations from the natural black-hole shape have to die off as one approaches the black hole limit under the physical principle that curvatures are bounded. 
\end{abstract}


\maketitle
 
\tableofcontents

\section{Introduction}
\label{Sec:Introduction}

No-hair theorems are one of the key results in General Relativity (GR), see e.g.~\cite{Chrusciel2012} for a comprehensive review. These theorems show that the only regular, asymptotically flat, vacuum, and stationary black-hole   configurations in GR are, under some additional assumptions, described by the two-parameter Kerr family of black holes. These theorems suggest that any dynamical process leading to the formation of a black hole, such as the gravitational collapse of a star to form a black hole, eventually settles to a member of the Kerr family. However, some shortcomings are found in applying these theorems to realistic situations. 

The standard theorems use as hypotheses local requirements, for instance, regularity of the horizon surface or the Einstein equations, together with nonlocal properties such as asymptotic flatness or constraints on the topology of constant redshift surfaces. However, it does not seem to be clear what is the role that each of these assumptions plays when considered separately. For example, can one distort a horizon while maintaining its regularity? Is there only some kind of specific distortions that can be applied to black holes? These and other questions acquire crucial relevance when considering departures from the idealized isolated black hole model. 

On the one hand, black holes are rarely isolated; they typically exist in (quasi-)equilibrium states surrounded by some form of matter or external gravitational fields. Although it is widely believed that such black holes remain ``hairless", it is not always crystal clear what is understood by that, namely, concerning the question of whether only a two-parameter family of black holes would be compatible with the structure of GR. 
To the best of our knowledge, the only results reported in that direction are those of G\"urlebeck~\cite{Gurlebeck2012,Gurlebeck2015}. G\"urlebeck showed, in the static and axisymmetric case, that the Weyl multipole moments that characterize a spacetime close to infinity can be expressed as a sum of the contributions from black hole regions and regions with a non-vanishing energy-momentum tensor. The contributions from black holes are those associated with the Schwarzschild black holes. That is, a distorted black hole does not modify the contribution to the multipolar structure of the external gravitational field in which it is embedded. However, for instance, there is no statement of a no-hair theorem asserting the existence of a single family of distorted black hole solutions given an external gravitational background.  

On the other hand, the original formulation of the no-hair theorem requires the presence of an event horizon, specifically an infinite-redshift surface. However, recent results~\cite{Barcelo2019,Raposo2019} show that demanding the curvature at a surface of arbitrarily high redshift to be bounded and imposing vacuum Einstein equations up to spatial infinity places strong bounds on the multipolar structure of   spacetime. Furthermore, the multipole structure of the configuration approaches that of a black hole as the redshift of the surface parametrically goes to infinity. Hence, these results indicate that the requirement of an infinite-redshift surface can be relaxed without altering the overall physical picture. All our arguments apply to the exterior region of black holes but not their interior region, which is also the case for the classic no-hair theorems.

In this paper, we delve into these two aspects of the no-hair theorems, carefully analyzing the static and axisymmetric case. This restricted case allows for a clear conceptual treatment. The static nonaxisymmetric and the stationary axisymmetric cases exhibit additional complications that deserve separate in-depth analyses beyond this paper. Thus, in this paper, we first sharpen the claim made in~\cite{Gurlebeck2015} about the no-hair theorem for black holes in external gravitational fields. We show that, given an external gravitational field, there is at most a one-parameter family of black hole geometries compatible with it, provided an equilibrium condition is fulfilled. We show that the parameter describing this family can be identified with the mass of the black hole in the limit in which the external gravitational field is turned off. Second, we illustrate in a clear way the difference between deforming black holes from the ``inside'' and from the ``outside''. As we will explain later, we will call deformations from the inside those that imply a modification of the source itself generating the black hole. In the axisymmetric problem and in Weyl coordinates this source is distributional (i.e., delta like) and it is localized at the horizon itself (as we will see, this source does not translate into a real GR stress-energy source). In contrast, sources located outside that region correspond to deformations from the ``outside'', and we identify them with external gravitational fields. Whereas the former give rise to arbitrarily large curvatures in the surface as the redshift is increased, the latter give contributions that remain finite. Finally, we take the opportunity to clarify some misconceptions regarding the deformability of black hole horizons and the meaning of Love numbers.

Here is an outline of the article. In Section~\ref{Sec:IsraelRev} we review Israel's theorem and revisit the different assumptions made in its derivation, with special emphasis on disentangling local and global aspects of the theorem. In Section~\ref{Sec:NoHair} we present the main tools and results that will be used in the article. In Subsec.~\ref{Subsec:Putative} we discuss the different vacuum solutions that exhibit an infinite redshift at some region and argue that only one of them leads to a smooth event horizon. In Subsecs.~\ref{Subsec:InsideDefs}
and~\ref{Subsec:OutsideDefs} we illustrate the difference between deforming a horizon from inside (some source or modification of GR inside the high-redshift surface) and outside (external gravitational fields). In Section~\ref{Sec:NoHair_External} we state a no-hair result for static and axisymmetric black holes in external gravitational fields based on the former analysis. In Section~\ref{Sec:NoHair_ECOs} we apply this machinery to the set-up in which there exists a maximum redshift surface with bounded curvatures and find bounds to the deviation of the multipole structure with respect to the case in which the redshift goes to infinity (the black hole limit). Finally, we conclude in Section~\ref{Sec:Conclusions} with a summary of our work and comment on some aspects of no-hair theorems in astrophysical contexts and the interplay between our results and the vanishing of Love numbers. In Apps.~\ref{App:Topology_Horizons},~\ref{App:Rings} and~\ref{App:Curzon}, we have collected some technical derivations and statements of some known results that have been skipped in the main text for clarity of the presentation.

 \paragraph*{\textbf{Notation and conventions.}}

We use the signature $(-,+,+,+)$  for the spacetime metric and we work in geometrized units in which $c=G_N=1$. For the curvature tensors we use the conventions in the book of~\cite{Wald1984} {\it i.e.}, $[\nabla_A, \nabla_B] V^C =: - R_{ABD}{}^C V^D$, $R_{AB}:=R_{A C B}{}^C$. We use capital latin indices from the beginning of the alphabet as indices in the 4-dimensional manifold $(A, B...)$, lower case indices from the middle of the alphabet $(i,j...)$ for indices within the 3-dimensional submanifolds defined by constant time slices, and lower case indices from the beginning of the alphabet $(a,b...)$ for indices in 2-dimensional surfaces within these 3-dimensional submanifolds. The symbol $\Delta$ used at several places in the draft denotes a difference and not the Laplacian, which is always denoted by $\nabla^2$.

\section{Revisiting Israel's theorem: local vs global aspects}
\label{Sec:IsraelRev}
Let us review Israel's theorem with the aim of disentangling local and global aspects of the theorem, which will be essential for formulating the extensions presented afterwards. The starting point of the discussion is a general static metric which can be written in the following form~\cite{Israel1967}:
\begin{align}
    ds^2 = - \mathscr{V}^2(x) dt^2 + g_{ij} (x) dx^i dx^j,
\end{align}
where $\mathscr{V}(x)$ represents a potential function, and $g_{ij}$ is an arbitrary static 3-dimensional metric. This parametrization can be used to describe any horizonless configuration, with $\mathscr{V}\neq 0$, or in the case of having horizons, i.e. surfaces at which $\mathscr{V} \to 0^+$, the geometry exterior to them. From a pure geometrical  (i.e. kinematical) perspective, it is interesting to recall that a geometry can have horizons with any shape and topology. For example, one can build a black hole with a horizon 2-surface with genus-2 topology. Hawking theorem on the topology of black holes in 4D makes the additional dynamical assumption that the matter content must satisfy the dominant energy condition~\cite{HawkingEllis1973}. In fact, this theorem shows that the topology of a horizon must be spherical or exceptionally toroidal. See App.~\ref{App:Topology_Horizons} for further details on the specific axisymmetric case.

Now, the vacuum Einstein equations for the 3-dimensional metric can be written in the following form:
\begin{align}
    & {}^{3}R = 0, \\
    & {}^{3}R_{ij} + \mathscr{V}^{-1} \nabla^{(g)}_i \nabla^{(g)}_j \mathscr{V} = 0, 
\end{align}
where ${}^{3}R_{ij}$ and ${}^{3}R$ represent the Ricci tensor and the Ricci scalar of the metric $g_{ij}$ and $\nabla^{(g)}_i$ is the covariant derivative constructed with the Levi-Civita connection of $g_{ij}$. Combining both equations, we can replace the equation associated with the vanishing of the Ricci scalar ${}^{3}R$ with a Laplace equation for $\mathscr{V}$,
\begin{align}
    \nabla^2_{(g)} \mathscr{V} = 0. 
\end{align}
Israel's theorem~\cite{Israel1967} imposes the following conditions: 
\begin{itemize}
    \item Take $\Sigma$ to be any $t=\text{constant}$ spatial hypersurface that is regular, empty, noncompact, and asymptotically flat, namely that there exists a set of coordinates (that are asymptotically identified with the coordinates $x^i$) in which:
    \begin{align*}
        & g_{ij} = \delta_{ij} + \order{r^{-1}}, \qquad \partial_{k} g_{ij} = \order{r^{-2}}, \\
        & \mathscr{V} = 1 - m/r + \eta (r), \qquad m = \text{constant}, \\
        & \eta = \order{r^{-2}}, \qquad \partial_i \eta = \order{r^{-3}}, \qquad \partial_i \partial_j \eta = \order{r^{-4}}, 
    \end{align*}
    when $ r = \left( \delta_{ij} x^i x^j \right)^{1/2} \rightarrow \infty $.
    \item The equipotential surfaces $\mathscr{V} = \text{constant} >0,$ $t = \text{constant}$ are regular, simply connected, and closed 2-spheres. 
    \item The Kretschmann scalar $\mathcal{K} = R_{ABCD} R^{ABCD}$ is bounded everywhere on $\Sigma$. 
    \item If $ \mathscr{V} $ has a vanishing lower bound on $\Sigma$, the two-surfaces  $\mathscr{V}=c \in \mathbb{R}$ approach a limit as $c \rightarrow 0^{+}$, corresponding to a closed regular 2-space of finite area.  
\end{itemize}
Israel showed that the only static metric that obeys all these conditions is the Schwarzschild metric, which also turns out to be spherically symmetric. The core of the proof is a sort of shooting method in which the equations above are integrated from the asymptotic $\mathscr{V} = 1$ surface to the $ \mathscr{V} = 0^{+}$ surface. Then, Israel shows that the Schwarzschild solution is the only one satisfying all the assumptions and resulting in a regular geometry at $\mathscr{V}=0^{+}$. In order to implement this procedure, the first step is to introduce a set of adapted coordinates on the spatial hypersurfaces, which correspond to the coordinate $\mathscr{V}$ and a set of angular-like coordinates $(\theta^1,\theta^2)$ on the surfaces $\mathscr{V}=\text{constant}$. 

The first hypothesis sets the problem in the sense that one considers a vacuum spacetime and ensures that the spacelike slices are noncompact and have an asymptotic infinity. Furthermore, the asymptotically flat behavior can be seen as providing the initial data for the inward integration. The second assumption, in combination with the emptiness assumption, is required to ensure that there is only ``one single spheroidal central object'' acting as a source of the gravitational field. Relaxing these assumptions (vacuum and single central spheroid) is one of the aims of this paper.

Finally, the last two assumptions ensure that the geometry is regular at the surface $\mathscr{V} = 0^{+}$, i.e., that the spacetime displays a regular Killing horizon (which in this case coincides with the event horizon). In this way, the shooting procedure singles out the Schwarzschild geometry, since any deviation from the Schwarzschild metric would lead to a singular geometry (with a singular Kretschmann scalar) on the event horizon, thus concluding the uniqueness of the solution. As a side note, it is puzzling to realize that the theorem specifically avoids the analysis of the strange case in which a putative horizon of infinite area forms (more on this later on).

From this result, we could draw  too quickly the conclusion that any distortion of an event horizon away from spherical symmetry leads to a singular behavior. This misunderstanding may be reinforced due to the reported vanishing of the Love numbers of black holes (see~\cite{Binnington2009,Poisson:2020vap} for an introduction and~\cite{Damour2009,Damour2009b,Kol2011,Charalambous:2021mea} for the computation for static black holes) being often interpreted as the impossibility of deforming an event horizon. However, in this paper we will clearly show that we can separate the deformations of an event horizon from spherical symmetry in terms of the ``agent'' that generates the deformation of the horizon: whether it is an intrinsic deformation (due to the structure of the object itself), or a deformation due to an external gravitational field (due to the presence of a disk of matter; the only external matter consistent with axisymmetry). Whereas the former kind of deformations does indeed lead to a singular geometry at $\mathscr{V} = 0^{+}$, the latter gives rise to a perfectly smooth event horizon which departs from spherical symmetry. 

Actually, let us show that the vacuum Einstein equations are compatible with any geometry induced at the horizon, as long as it is topologically a 2-sphere (times an open interval representing time). We can show this through a constructive procedure following the approach in~\cite{Frolov1985} that we sketch in here. If an analytic static metric exists in an open neighborhood outside a regular topologically-spherical horizon (note that we are excluding more exotic mathematical situations such as the interesting nowhere-differentiable horizons discussed in~\cite{Chrusciel1996}), then one can conveniently choose adapted coordinates $(t,\mathscr{V},\vec{\theta})$ so that the metric reads:
\begin{align}
    ds^2 = -\mathscr{V}^2 dt^2 + \frac{d\mathscr{V}^2}{\kappa^2 (\mathscr{V}, \vec{\theta})} + h_{ab} (\mathscr{V}, \vec{\theta}) d\theta^a d \theta^b,
\end{align}
with the horizon located at $\mathscr{V} \rightarrow 0^{+}$. Then, it is possible to write the Einstein vacuum equations in a neighborhood of $\mathscr{V} = 0$ for $\kappa$ and $h_{ab}$ and solve them for a series expansion of the form:
\begin{align}
    \kappa = \sum_{n = 0}^{\infty} \kappa^{(n)} ( \vec{\theta} ) \mathscr{V}^n, \qquad h_{ab} = \sum_{n =0 }^{\infty} h_{ab}^{(n)} (  \vec{\theta} ) \mathscr{V}^n.
\end{align}
One can show that the consistency of this procedure requires $\kappa^{(0)}$ to be a constant (in agreement with the zeroth law of black hole dynamics~\cite{Bardeen1973}), which corresponds to a choice of units. Apart from this condition, the Cauchy-Kovalevskaya theorem~\cite{Evans2010} ensures that there always exists a unique and analytical local solution for every analytic tensor $h_{ab}^{(0)}(\vec{\theta})$. Hence, it is possible to have a local black hole with any kind of horizon shape. Then, one can continue integrating outwards starting from one of these arbitrarily shaped horizons. However, Israel's theorem tells us that only for the purely spherically symmetric shape, integrating outwards using the Einstein vacuum equations leads to an asymptotically flat spacetime. We will show this explicitly in the next section, in which we restrict our attention to the case of static and axisymmetric solutions, where it is possible to provide sharper and clearer results.

Let us make two final remarks before discussing static and axisymmetric solutions in the next section. First, a static horizon is only compatible with it being in vacuum or immersed in matter that violates energy conditions. Hence, if we impose that the dominant energy condition is obeyed on the horizon, there should be a neighbourhood of the horizon devoid of matter~\cite{Wald1984}. Second, it is convenient to rewrite the line element for a static metric in a form which is closer to the line element that we will use in the static and axisymmetric case:
\begin{align}
    ds^2 = - e^{2U} dt^2 + e^{-2U} \gamma_{ij} dx^i dx^j, 
    \label{Eq:gamma_metric}
\end{align}
where $\gamma_{ij}$ is an Euclidean 3-dimensional metric. This form is related to the previous line element through the mapping $e^{2U} = \mathscr{V}^2$ and $g_{ij} = e^{-2 U } \gamma_{ij}$. The Einstein equations in vacuum for this form of the metric can be written as
\begin{align}
    &  R_{ij} \left( \gamma \right) + 2 \nabla^{(\gamma)}_{i} U \nabla^{(\gamma)}_j U = 0, \\
    & \nabla^2_{(\gamma)} U = 0,
\end{align}
where $R_{ij} \left( \gamma \right)$ represents the Ricci tensor of the metric $\gamma_{ij}$ and $\nabla_{(\gamma)}^2$ represents the Laplacian with respect to the metric $\gamma$ itself. A putative horizon in these coordinates would be located at $U \rightarrow - \infty$.   
\section{Static and axisymmetric geometries}
\label{Sec:NoHair}
Let us restrict the discussion to static and axisymmetric geometries since the simplicity of this setup allows for an exhaustive explicit analysis. Furthermore, we will assume that the topology of the horizon is a sphere (in 4 dimensions it can actually be only a sphere or a torus, see App.~\ref{App:Topology_Horizons}, where we also analyze the latter case). These metrics can be written in the Weyl form, that is well adapted for this set of isometries: 
\begin{align}
    ds^2 = - e^{2 U} dt^2 + e^{-2 U} \left[ e^{2V} \left( dr^2 + dz^2 \right) +r^2 d \varphi^2 \right],
\label{Eq:Metric}
\end{align}
where $U = U(r,z)$ and $V=V(r,z)$ are functions that depend only on $r$ and $z$. The vector generating the time translations is $\bm{k} = \partial_t$ and the vector generating the axisymmetric transformations is $\bm{m} = \partial_{\varphi}$. This is a particular form of the generic static metric presented in Eq.~\ref{Eq:gamma_metric} with the metric $\gamma_{ij}$ expressed in coordinates adapted to staticity and axisymmetry. More explicitly, we can write in these coordinates:
\begin{align}
    \gamma_{ij} dx^i dx^j = e^{2 V} \left( dr^2 + dz^2 \right) + r^2 d \varphi^2.
\end{align}
This form can always be achieved by means of coordinate transformations since the Lie algebra spanned by these two vectors is trivial, $[\bm{k},\bm{m}] = 0$~\cite{Carter1970}. 

In the presence of matter ($T^{\mu}_{\ \nu} \neq 0$) Einstein equations are given by:
\begin{align}
8 \pi  T^t_{\ t} & = e^{2 U - 2V} \left[ -  2 \nabla^2 U + \nabla^2 V - \frac{1}{r} \partial_r V + \left( \partial_r U \right)^2 + \left( \partial_r U \right)^2 \right], \nonumber \\
8 \pi T^{r}_{\ r} & = e^{2U - 2V}  \left[- \left( \partial_r U \right)^2 + \frac{1}{r} \partial_r V +\left( \partial_z U \right)^2  \right], \nonumber \\
8 \pi T^{r}_{\ z} & = e^{2U - 2V} \left[ -2 \partial_{r} U \partial_{z} U + \frac{1}{r} \partial_{z} V \right], \nonumber \\
8 \pi T^{z}_{\ z} & = e^{2U - 2V} \left[ \left( \partial_r U \right)^2 - \frac{1}{r} \partial_r V -\left( \partial_z U \right)^2  \right], \nonumber \\
8 \pi  T^{\varphi}_{\ \varphi} & = e^{2 U- 2 V} \left[ \nabla^2 V - \frac{1}{r} \partial_r V + \left( \partial_r U \right)^2 + \left( \partial_z U \right)^2 \right].
\label{Eq:Einsteineqs}
\end{align}
In these equations $\nabla$ represents the covariant derivative associated with a fiduciary Euclidean 3-space, expressed in cylindrical coordinates $(r,z,\varphi)$. As we will see, the appearance of this fiduciary Euclidean 3-space is a formal advantage that permits a complete control of the static axisymmetric case as opposed to the general static case. 

We can subtract the first equation from the last one and manipulate it to find:
\begin{align}
    \nabla^2 U = 4 \pi e^{2V- 2U} \left( - T^{t}_{\ t} + T^{\varphi}_{\ \varphi}  \right),
    \label{Eq:NewtonAnalogue}
\end{align}
which is similar to an effective Poisson equation in a flat 3-space, although the source depends nonlinearly on $U$ and $V$. In empty regions of spacetime we simply have  
$\nabla^2 U=0$; in this case, we will call finding $U(r,z)$ the \emph{associated Laplacian problem} (ALP).

In Newtonian gravity, given some sources, one only needs to solve Poisson equation on a flat 3D space to obtain the gravitational potential $U$. Any source is compatible with the equation, for example, two masses at rest at a fixed distance. Poisson equation by itself does not tell you whether a configuration of sources that interact only gravitationally will be consistent or not with staticity. However, GR is a constrained system. One cannot give arbitrary conditions but only those compatible with the constraints. In this specific case this general issue appears in the solutions of the equation for $V$. As we will see, even if $U$ is perfectly regular it might happen that the associated $V$ exhibits irregularities. These irregularities inform us about some inconsistency that the selected source might have. In particular, in the absence of matter, the function $V$ must vanish on the $z$-axis, i.e. $V(0,z)=0$, to avoid the presence of conical singularities. This ensures that the length of a proper orbit of the axial Killing vector $\bm m$ that passes through a point $q$ located at a proper distance $d$ from the axis (fixed points of $\bm m$) is $2 \pi d$ at leading order (see~\cite{Mars1992} for a rigorous demonstration).

\begin{figure}
\begin{center}
\includegraphics[width=0.25 \textwidth]{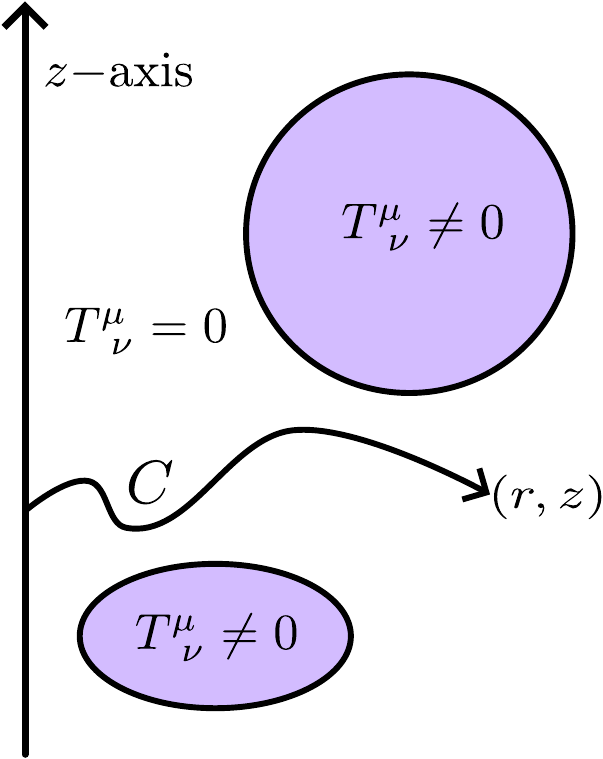}
\caption{Schematic representation of the problem. The $z$-axis is depicted as a black solid line. We illustrate the presence of matter in some compact regions that are colored. Notice that those regions would correspond to topological tori once we consider the $\varphi$ direction too. Furthermore, we depict a generic curve $C$ that begins in the $z$-axis and ends up in the point of coordinates $(r,z)$. The integral of $\boldsymbol{\omega}$ along $C$ would give us the function $V(r,z)$.}
\label{Fig:Sketch}
\end{center}
\end{figure}

To delve into this problem, we first realize that, in regions devoid of matter, the second and third equations in~\eqref{Eq:Einsteineqs} can be rewritten as
\begin{align}
    & \partial_r V = r \left[ (\partial_r U)^2 - ( \partial_z U)^2 \right], \label{Eq:DrV}\\
    & \partial_z V = 2 r \partial_r U \partial_z U, \label{Eq:DzV}\\
    & \nabla^2 U = 0, 
\end{align}
where the last equation is precisely the ALP for $U$. It is convenient now to introduce the one-form (in the $rz$ plane)
\begin{align}
    \boldsymbol{\omega} := r \left[ (\partial_r U)^2 - ( \partial_z U)^2  \right] dr + 2 r \partial_r U \partial_z U dz.
\end{align}
This form is defined wherever the coordinates are valid, even if $\nabla^2 U\neq0$. If we take the exterior derivative of this form, we find that it is proportional to the Laplacian of $U$, namely we find 
\begin{align}
    d \boldsymbol{\omega} = -2 r \left( \nabla^2 U \right) \left(\partial_z U \right) dr \wedge dz.
\end{align}
Thus, for every simply connected open set of the spacetime which is in vacuum, we can write $\boldsymbol \omega$ as the differential of the scalar function $V$. Thus, in vacuum open sets, once the ALP for $U$ is solved, $V$ can be determined simply by performing an integral of $\boldsymbol\omega$, which  is defined in terms of $U$ through Eqs.~\eqref{Eq:DrV}-\eqref{Eq:DzV}, as
\begin{align}
   V (r,z) = V(r_0,z_0) +  \int_{C} \boldsymbol{\omega},
\end{align}
with $C$ being any curve that ends in the point with coordinates $(r,z)$ and $V(r_0,z_0)$ represents a boundary condition for the function $V$ that is required to solve the full problem. Given the regularity condition that we have to impose on the $z$-axis, we choose $r_0=0$ so that $V(0,z_0) = 0$. The curve $C$ is any curve that starts at the $z$-axis  such that it avoids the regions in which there is some matter content for this to hold. In fact, in this work we will be always focusing on the vacuum regions. See Fig.~\ref{Fig:Sketch} for a pictorial sketch of the setup. 
The simplest curve that one can take is precisely a curve that is orthogonal to the $z$-axis which leads to the following expression for $V(r,z)$:
\begin{align}
    V(r,z) = \int^{r}_{0} dr' r' \left[ \left(\partial_{r'} U(r',z)\right)^2 - \left(\partial_z U(r',z) \right)^2 \right].
    \label{Eq:V_cuadrature}
\end{align}
In addition, asymptotic flatness requires that $V(r,z)$ vanishes also at infinity. This cannot be ensured automatically through an additional boundary condition, since the equations for $V(r,z)$ are of first order and we can only impose one boundary condition. Since we have decided that $V(0,z)=0$, the vanishing at infinity would require that $U(r,z)$ obeys the following non-local constraint
\begin{align}
    \int_0^{\infty} dr r \left[ \left(\partial_{r} U(r,z)\right)^2 - \left(\partial_z U(r,z) \right)^2 \right]= 0. 
    \label{eq:equilibriumzeroinfty}
\end{align}
By inspecting some examples, such as a spherically symmetric star surrounded by a ring (a Saturn-like configuration), we realize that whenever the matter content linked to the ALP is not in equilibrium, this integral does not vanish. For example, for a ring located in the equatorial plane, which divides the star in two equal half-spheres, the integral straightforwardly vanishes. When the ring is not on the equatorial plane, the integral gives a non-vanishing result. We will illustrate this explicitly in Subsec.~\ref{Subsec:OutsideDefs} with a black hole solution surrounded by a matter ring. 

Let us  now  establish a firm  relation between regularity at the axis $V(0,z)=0$, asymptotic flatness $V(\infty,z)=0$, the condition in Eq.~\eqref{eq:equilibriumzeroinfty}, and the notion that the effective matter content in the ALP is in equilibrium. With this aim, let us first prove that Eq.~\eqref{eq:equilibriumzeroinfty} is equivalent to
\begin{equation}
\oint_\gamma\boldsymbol\omega=0
\label{eq:equilibriumcond}   
\end{equation}
for any closed curve $\gamma$ fully contained in the vacuum region.
\begin{figure}
\begin{center}
\includegraphics[width=0.75 \textwidth]{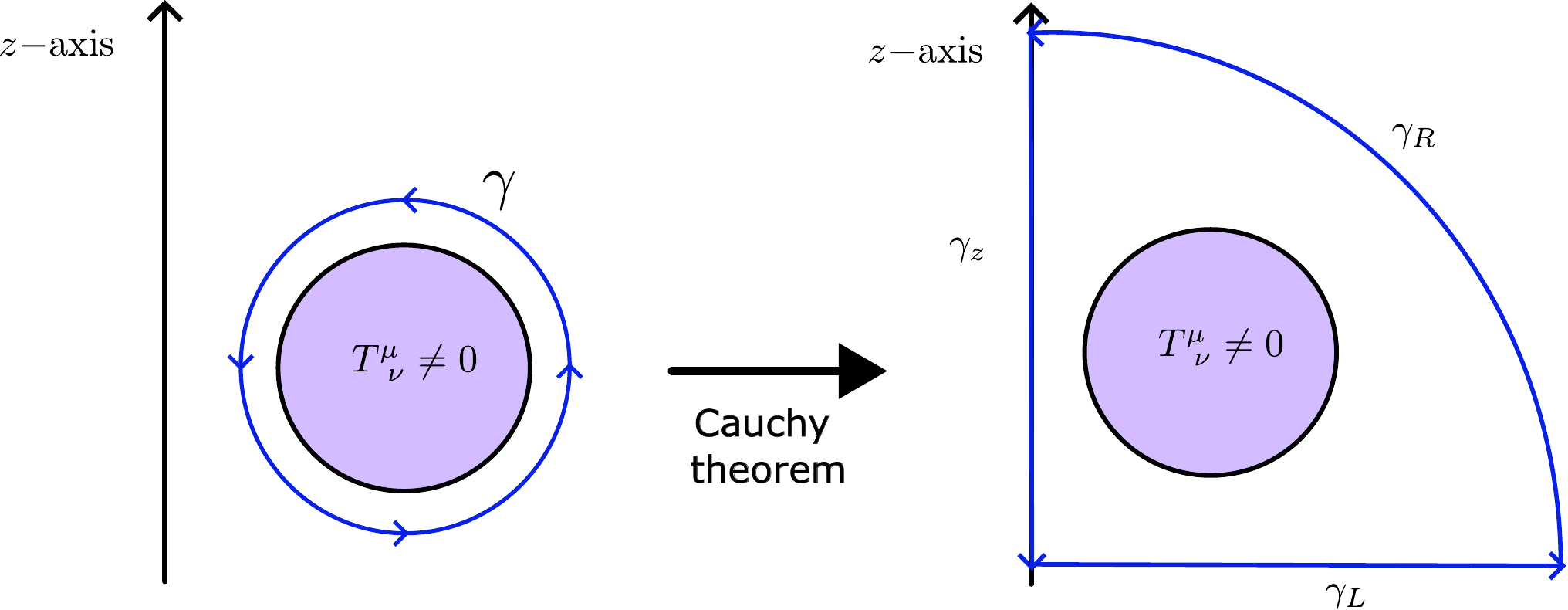}
\caption{Schematic representation of the argument to show the equivalence between Eq.~\eqref{eq:equilibriumzeroinfty} and Eq.~\eqref{eq:equilibriumcond}. The $\gamma$ contour can be deformed into the $\gamma_z+\gamma_L+\gamma_R$ due to Cauchy theorem. The integral along $\gamma_z$ is shown to vanish at the axis whereas the integral along $\gamma_L$ which is precisely the integral in Eq.~\eqref{eq:equilibriumzeroinfty}.}
\label{Fig:Sketch2}
\end{center}
\end{figure} 
A closed contour $\gamma$ as the one depicted in the left of Fig.~\ref{Fig:Sketch2} can be deformed to the contour $\gamma_z+\gamma_L+\gamma_R$ in the right without changing the value of the integral (Cauchy's theorem):
\begin{align}
\oint_\gamma\boldsymbol\omega=\int_{\gamma_L}\boldsymbol\omega+\int_{\gamma_R}\boldsymbol\omega+\int_{\gamma_z}\boldsymbol\omega.
\end{align}
In the limit $R\to\infty$, the first term in the right hand side is simply the integral in Eq.~\eqref{eq:equilibriumzeroinfty}.
The second term can be obtained   from a series expansion in $1/R$ (multipolar expansion) of $U$,
whose leading order is $1/R$. Therefore, 
\begin{align}
    \int_{\gamma_{R}} \boldsymbol{\omega} \sim-\frac{1}{R^2}+  \order{\frac{1}{R^4}},
\end{align}
which vanishes in the limit $R \rightarrow \infty$. 
Finally, the last integral is equal to $V(0,z_0)-V(0,R)$; Eq. \eqref{Eq:DzV} implies that $V$ is constant at the axis and therefore this term also vanishes. Consequently, $\oint_\gamma\boldsymbol\omega=0$ if and only if Eq.~\eqref{eq:equilibriumzeroinfty} holds as we wanted to show.

Now we can use Stoke's theorem to rewrite the condition $\oint_\gamma\boldsymbol\omega=0$  as a surface integral over the surface $S$ enclosed by $\gamma$ (in the $rz$-plane)
\begin{align}
   \oint_\gamma\boldsymbol\omega = \int_{S} dz dr r \partial_z U\nabla^2 U=0.
    \label{Eq:Equilibrium_Condition}
\end{align}
In the absence of non-empty regions, it vanishes identically because $\nabla^2 U=0$. However, if the closed curve $\gamma$ encloses a surface $S$ with some matter content, it is not straightforward that it vanishes since the integrand is nonvanishing. In the Newtonian limit, the vanishing of the projection of the net force on the $z$-axis $ F_{z}=-\partial_z U $ is an equilibrium condition. In the fully relativistic setup, it translates into the vanishing of the integrals in Eq.~\eqref{Eq:Equilibrium_Condition}, which is a necessary and sufficient condition for the compatibility of regularity and asymptotic flatness as we have seen, where the role of the (Newtonian) matter density is now played by an effective density which is not the energy density alone as we see from Eq.~\eqref{Eq:NewtonAnalogue}. 
For this reason, we will call Eq.~\eqref{Eq:Equilibrium_Condition} the \emph{equilibrium condition.} 

The non-vanishing of the integral in Eq.~\eqref{Eq:Equilibrium_Condition} (i.e. a violation of the equilibrium condition) is translated to the fact that asymptotic flatness and regularity of the axis cannot hold simultaneously. This means that there has to be some matter either at the axis or at infinity that supports the configuration and is responsible for the staticity of a system which is not in equilibrium. In that sense, the non-vanishing of $V$ at the $z$-axis would correspond to a cosmic-string-like behavior on the axis of the energy-momentum tensor and the non-vanishing of $V$ at infinity would correspond to some matter content located at infinity.

\subsection{Putative black hole spacetimes}
\label{Subsec:Putative}

For static metrics, an event horizon is a Killing horizon~\cite{HawkingEllis1973} and hence, at a horizon, the redshift function $e^{U}$ should vanish. This means that the function $U$ for which we need to solve an ALP, needs to go to $-\infty$ at any putative horizon. Thus the horizon would not be covered by these coordinates. The regions at which the function $U$ goes to $-\infty$ must correspond to those in which there is some kind of singular ($\delta$-like distributional) source. However, we highlight that a proper horizon itself would be in vacuum, and these sources would become an artifact of the ALP. Here, we focus exclusively on smooth horizons, specifically at least $\mathcal{C}^2$, to ensure the Einstein equations remain well-defined. However, beyond this framework, exotic horizons that are nowhere differentiable can be constructed~\cite{Chrusciel1996}, though we do not consider such cases in our analysis.

We focus on asymptotically flat spacetimes with matter (either corresponding to real matter or to distributional matter that represents the location of a horizon) restricted to live in a compact region of spacetime. In such setup, the potential sources that might lead to a blow-up of $U$ are necessarily objects of codimension 3 or 2, i.e., point-like contributions or rod-like contributions. Other extended objects of codimension 1 or 0 cannot generate divergences in $U$ if they are located in a compact region. 

Given the axisymmetric setup that we are considering, this translates into the fact that all the potential black hole sources need to be point-like and rod-like distributions located in the symmetry axis and any number of infinitely thin rings around the symmetry axis. Any solution of the ALP involving a divergence in $U$ associated with these elements is in principle a potential black hole candidate. At this stage, we will call them ``putative black holes'' since it is not clear whether these divergences can be identified with regular or singular horizons until further analysis. 

As the ALP is a linear equation, we can always separate the effects of the different putative black holes present in the system. Thus, let us discuss first the elementary cases in which there is a just one of these singular behaviors, namely a point-like distribution at the axis, a ring-like distribution around the symmetry axis, or a constant density rod-like distribution sitting on the axis. One could think of a fourth kind of distribution, corresponding to performing an identification of the coordinates in the symmetry axis lying at the ends of the rod, leading to a toroidal black hole. However, these black holes require matter that violates the energy conditions in the outside region, and we will not specifically focus on them (see Appendix~\ref{App:Topology_Horizons}, where we have included a description of these black holes for the sake of completeness).

\paragraph*{\textbf{Curzon's putative black holes.}}

\label{sec:curzon}

Let us first analyze a source term for the ALP which is a point-like distribution. The Curzon metric (and its higher multipole generalizations) correspond to a source term for the ALP which is precisely a point-like distribution. This can be a simple delta source (or even a more singular multipolar profile) located at a point on the axis. 

For the purpose of describing these metrics it is more convenient to change coordinates in the 3D space. Instead of using coordinates $(r,z,\varphi)$, we can use spherical coordinates $(R,\theta,\varphi)$. Thus we have $R = \sqrt{r^2 + z^2}$ and 
\begin{align}
    & r = R \sin \theta, \\
    & z = R \cos \theta.
\end{align}
Then, in these coordinates the metric in Eq.~\eqref{Eq:Metric} becomes:
\begin{align}
    ds^2 = - e^{2 U} dt^2 +e^{-2U} \left[ e^{2V} \left( dR^2 + R^2 d \theta ^2 \right) + R^2 \sin^2 \theta d \varphi^2 \right].
\label{Eq:Line-Element-v2}
\end{align}
Taking a monopolar solution for the ALP located at the origin, i.e. a source of the form $\rho(\boldsymbol{x}) = M\delta(\boldsymbol{x})$, yields
\begin{align}
    & U(R) = - \frac{M}{R}, \\
    & V(R, \theta) = - \frac{M^2 \sin^2 \theta}{R^2}, 
\end{align}
which is the so-called Curzon metric~\cite{Curzon1925}. We focus on the positive mass case here, since the negative mass case represents a naked singularity, with no horizon, and thus it is not of direct interest for us here. A more detailed analysis of the properties of the Curzon metric, including the negative mass case, is presented in Appendix~\ref{App:Curzon}. 

Notice that, although the redshift function does not depend on $\theta$, the whole metric is not spherically symmetric due to the functional dependence of $V$ on the angle $\theta$. This percolates to the Krestschman scalar which also depends on $\theta$:
\begin{align}
    \mathcal{K} = e^{2 M \left[M \sin ^2(\theta )-2 R\right]/{R^2}}
    \times {\mathcal{R}^{-4}(R,\theta;M)},
\end{align}
where $\mathcal{R}^4(R,\theta;M)$ is a function constructed from $M,R$ and $ \theta$, with dimensions of length to the fourth power:
\begin{align}
\mathcal{R}^4(R,\theta;M)= \frac{R^{10}}{8 M^2 \left[2 M^3 \sin ^2(\theta ) (M-3 R)-3 R^2 \left[M^2 (\cos (2 \theta )-3)+4 M R-2 R^2\right] \right]}.
\label{Eq:MathcalR}
\end{align}

Clearly it displays a directional behavior that is different from the Schwarzschild solution. For any value of $\theta \not \in \{0, \pi\}$ the Kretschmann scalar diverges as we approach $R \rightarrow 0$ since it has the behavior
\begin{equation}
\mathcal{K}_{\theta \not\in\{0, \pi\}} \sim e^{N M^2/R^2 + \cdots} \times \mathcal{R}^ {-4}(R,\theta;M),
\end{equation}
where $N$ is a positive number and we are omitting the subdominant term in the argument of the exponential. For $\theta \in\{0, \pi\}$ the Kretchmann has a different behavior and actually vanishes as we approach $R \rightarrow 0$ since in that case it exhibits a behavior 
\begin{equation}
    \mathcal{K}_{\theta \in\{0, \pi\}} \sim e^{-4 M/R} \times \mathcal{R}^{-4}(R,\theta;M),
\end{equation}
In both cases, the exponential behavior dominates either to make the Kretschman scalar blow up or vanish. The equipotential surfaces and the surfaces of constant Kretschmann scalar are depicted in Fig.~\ref{Fig:Curvatures_Redshift} (together with their Schwarzschild counterpart; see below). 
\begin{figure}[H]
\begin{center}
\includegraphics[width=0.45 \textwidth]{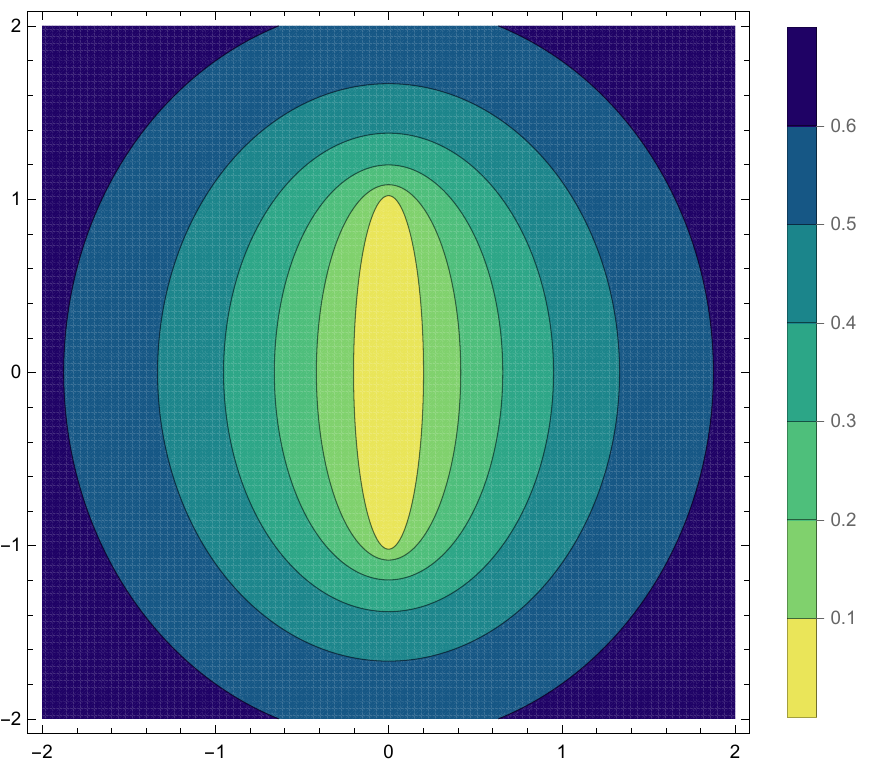}
\includegraphics[width=0.45 \textwidth]{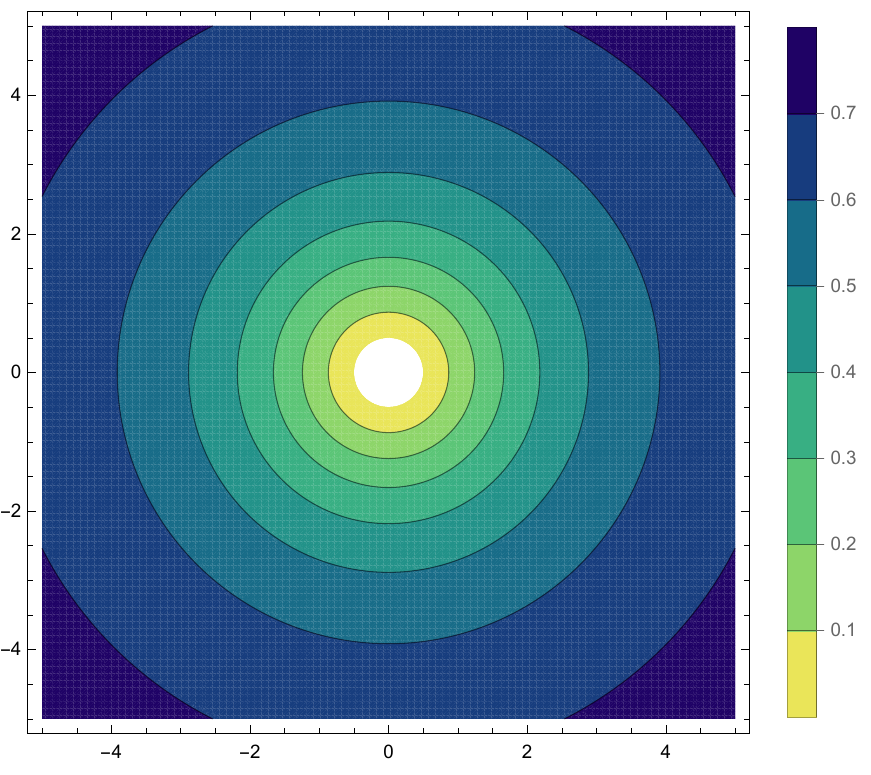}
\includegraphics[width=0.45 \textwidth]{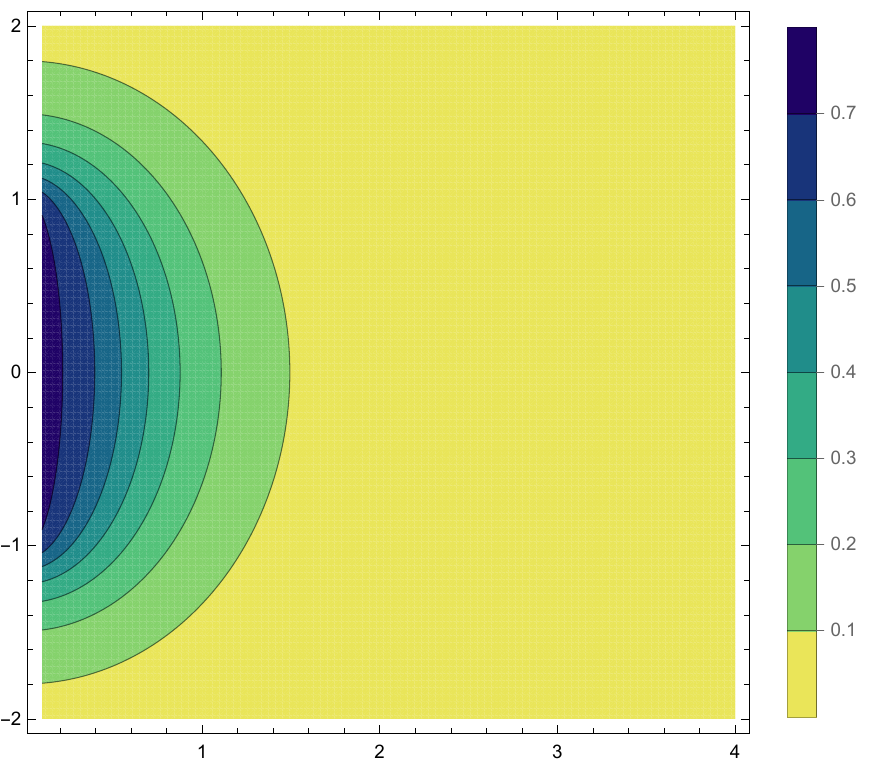}
\includegraphics[width=0.45 \textwidth]{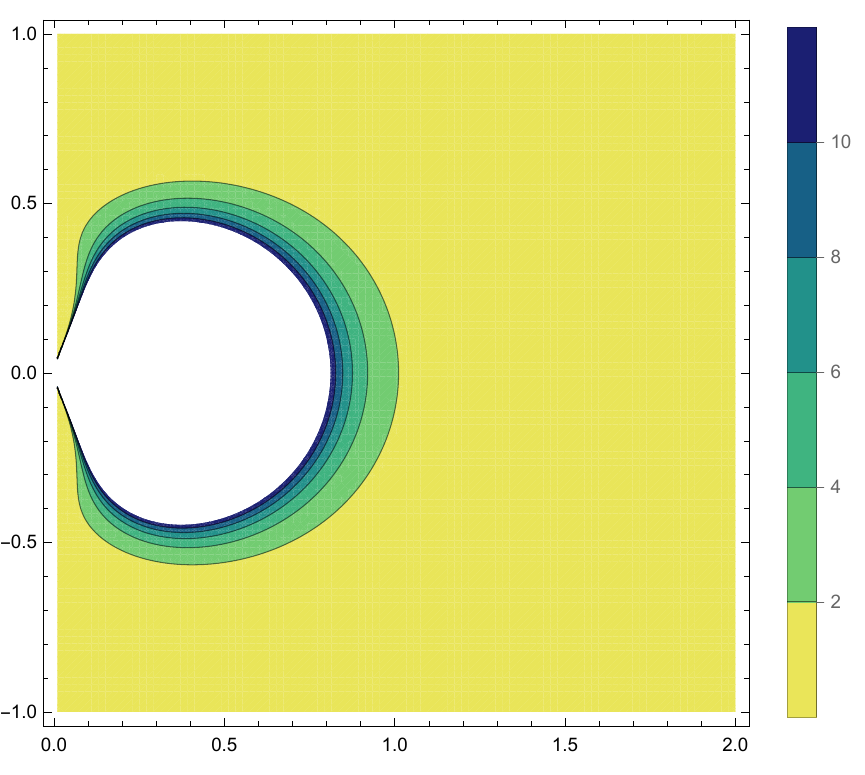}
\caption{We depict here in the upper panel the constant redshift surface in the $(r,z)$ coordinate for the Schwarzschild solution (left) and the Curzon solution (right), while we depict the Kretschmann scalar in the lower panel again for the Schwarzschild solution in the left and the Curzon solution to the right. We normalize the graphics so that the maximum is of $\order{1}$.}
\label{Fig:Curvatures_Redshift}
\end{center}
\end{figure} 
Another way of illustrating the non-spherically symmetric nature of the Curzon metric is by considering the induced metric on surfaces of constant redshift for the spatial sector of the metric:
\begin{align}
    ds_{U=\text{constant}}^2 = e^{-2U} \left[ e^{-2 U^2 \sin^2 \theta} d\theta^2 + \frac{M^2} {U^2} \sin^2 \theta d \varphi^2 \right],
\end{align}
which clearly lacks the spherical symmetry since its scalar curvature depends on $\theta$.

\begin{figure}[H]
\begin{center}
\includegraphics[width=0.75 \textwidth]{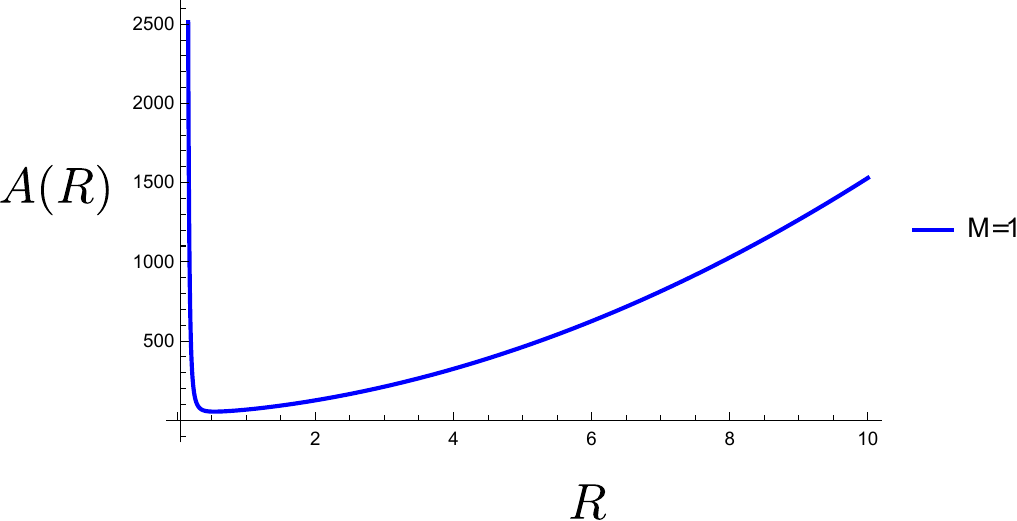}
\caption{We are depicting here the area function as a function of $R$ for $M=1$. The area decreases as we decrease $R$ until we reach a minimum $R \simeq 0.538905$ and it increases as we decrease the value of $R$ from that value on, see Appendix~\ref{App:Curzon} for further details.}
\label{Fig:AreaSingle}
\end{center}
\end{figure}

The Curzon metric exhibits another relevant feature. The area of the equipotential surfaces (constant $R$ surfaces) goes to infinity as $R \to 0$, see Fig.~\ref{Fig:AreaSingle} for a representation of the area of constant $t$ and $R$ surfaces as a function of the radius $R$. Its explicit analytic expression can be found in Appendix~\ref{App:Curzon} although it is not specially illuminating. This automatically excludes them from Israel's theorem scope, due to the last assumption of the theorem. Furthermore, it also shows that the putative horizon at $R \rightarrow 0$ is not regular in a different way. In fact, the directional behavior of the Kretschmann scalar, combined with the infinite extent of the constant $R$ region, suggests that the object, which appears point-like in these coordinates, actually represents an extended configuration.
In fact, the behavior of the geodesics in the Curzon metric was studied in~\cite{Morgan1973,Scott1985,Scott1985b}. Null geodesics approach the coordinate singularity $R=0$ in such a way that they enter parallel to the $z$-axis. Given that the surfaces $R = \text{constant}$ have an area that increases as we approach $R=0$, the surface $R=0$ is not really a point, but a whole plane of infinite extension. The geometry can be regarded as the end state of a very non-spherical anisotropic collapse which tends to compress all the matter into an infinitely diluted plane such that the matter content is moved away to infinity where a ring singularity is formed. Given that the plane is infinitely diluted, the curvature that is found by entering through the disk (which is what the null geodesics do in general, since they tend to enter towards the origin through the $z$-axis) is vanishing in an infinitely differentiable way, namely the $e^{-M/R}$ behavior (the archetypal example of a $C^{\infty}$ function which is not analytic). We can glue a flat spacetime to it which is what would be found by the light rays crossing the surface $R=0$ (which we emphasize is a plane, not a point). If we are to regard this as the end state of a collapse scenario, this is the only extension that we would do (as  the continuation to the interior in black hole situations). If we consider an eternal configuration, we would also do an extension backwards in time (continue the outgoing geodesics to the past) and glue another flat spacetime region. Further continuations are allowed involving the behavior of the other geodesics. Notice also that the extension is not unique, since the metric is not analytic, and hence there exists more (actually infinite) $C^{\infty}$ extensions of the metric across $R=0$. 

We can also consider the behavior of other singular pure multipole sources which are similar to the pure monopole behavior. The functions $U$ and $V$ in the metric for an $\ell$-order multipole configuration are \cite{Stephani2003}
\begin{align}
    & U^{(\ell)} = - \frac{M^{(\ell)}}{R^{\ell + 1}} P_{\ell} \left( \cos \theta \right), \nonumber\\ 
    & V^{(\ell)} = - \frac{(\ell + 1 ) \left[ M^{( \ell )} \right]^2}{2 R^{2 \ell + 2}} \left[ P_{\ell}^2 \left( \cos \theta \right) - P_{\ell+1}^2 \left( \cos \theta \right) \right].\label{Eq:U_Multipoles}
\end{align}
The Kretschmann scalar displays a behavior which is qualitatively similar to the monopole one as we approach $R\rightarrow 0$, although with a more convoluted directional structure. Instead of having only two directions along which the Kretschmann vanishes, there are now a set of directions $\theta \in \{ \theta_i \}_{i \in \mathcal{J}_{\ell}}$, with $\mathcal{J}_{\ell}$ a finite set for every $\ell$ along which the Kretschmann goes to zero. However, it is not easy to determine them in full generality and for every value of $\ell$. For any $\theta \neq \theta_i$, the leading behavior of the Kretschmann is 
\begin{align}
    \mathcal{K} \sim e^{N \frac{ \left[ M^{(\ell)} \right]^2}{R^{2 \ell + 2 }} + ... }\times \mathcal{R}^{-4}_{\ell} \left( R, \theta ; M^{(\ell)}\right), 
    \label{Eq:Kretschmann_Divergences}
\end{align}
where again $N$ is a given number and $\mathcal{R}^{-4}_{\ell} \left( R, \theta ; M^{(\ell)}\right)$ is a different polynomial that depends on $\ell$. For our purposes, it is enough to realize that for all the values of $\ell$, there exist directions for which the Kretchmann scalar becomes divergent as $R \rightarrow 0$. 

Exactly the same logic applies to the case in which we consider the superposition of different multipole configurations or even superpositions of the multipole configurations and the Schwarzschild solution, since the Kretschmann still contains these divergences that we have just discussed. 

All in all, this leads to the conclusion that the Curzon metric and its higher multipolar versions do not constitute proper black hole solutions since curvature singularities are found when approaching the vanishing redshift along some directions. They actually represent other kind of objects. For a further discussion of this issue, specially focused in the Curzon (monopolar) metric, see Appendix~\ref{App:Curzon}.

\paragraph*{\textbf{Ring singularities as putative black holes.}}

Consider the solution of the ALP  associated with an infinitely thin ring of matter located at $z=0$ and $r=r_{\rm ring}$, i.e. a distributional density  of the form $\rho(\boldsymbol{x}) = M_{\text{ring}}  \delta (z) \delta( r -r_{\rm ring} )/2 \pi$. This solution is often called the Bach-Weyl ring and is reviewed in detail in Appendix~\ref{App:Rings}. Axisymmetry implies that the ring has   constant linear density along its circumference. There are several pathological features of the geometry associated with this solution. 

First of all, as we approach the ring, the equipotential surfaces are toroids with smaller and smaller radii.
This implies that they do not satisfy the hypotheses of Israel's theorem. It is easy to see that the area of these equipotential toroids goes to zero as they approach the source since the vector generating axisymmetry $\boldsymbol{m}$ degenerates: The norm of $\boldsymbol{m}$ is given by $r e^{U}$, and $r$ remains bounded whereas $e^{U}$ vanishes as $r \rightarrow r_{\rm ring}$. Thus, the area of the toroids also vanishes since it is proportional to $2 \pi$ times the norm of $\boldsymbol{m}$ times the integral in the additional direction. This by itself tells us that they cannot approach a proper black hole horizon (of finite well-defined area) as it was noticed by Geroch and Hartle~\cite{Geroch1982}, which led them to exclude this possibility as potential black hole configurations in the static-axisymmetric setup. 
\begin{figure}
\begin{center}
\includegraphics[width=0.35 \textwidth]{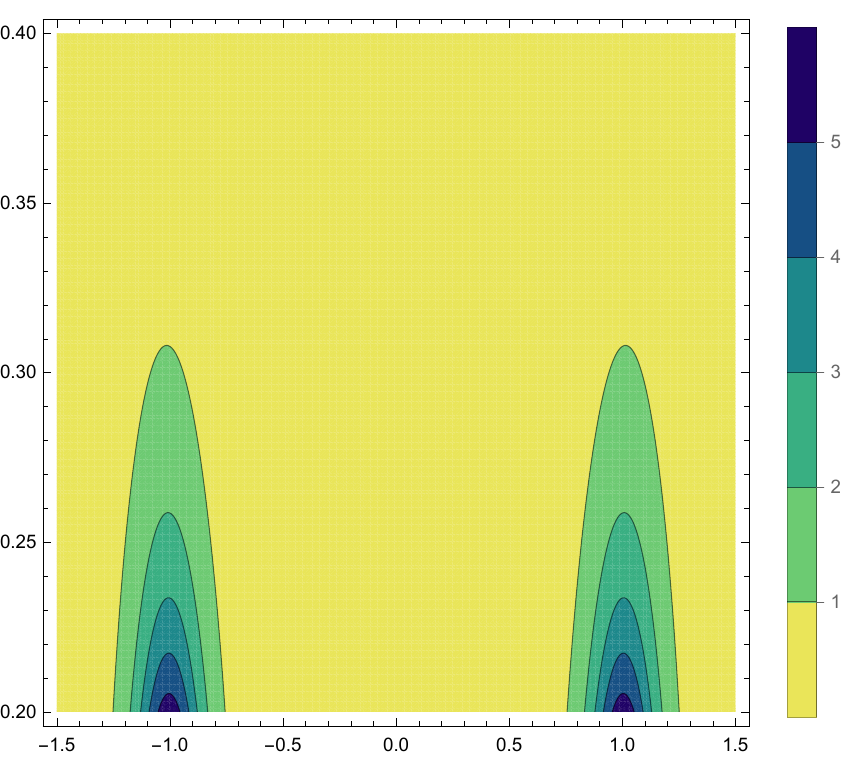}
\includegraphics[width=0.45 \textwidth]{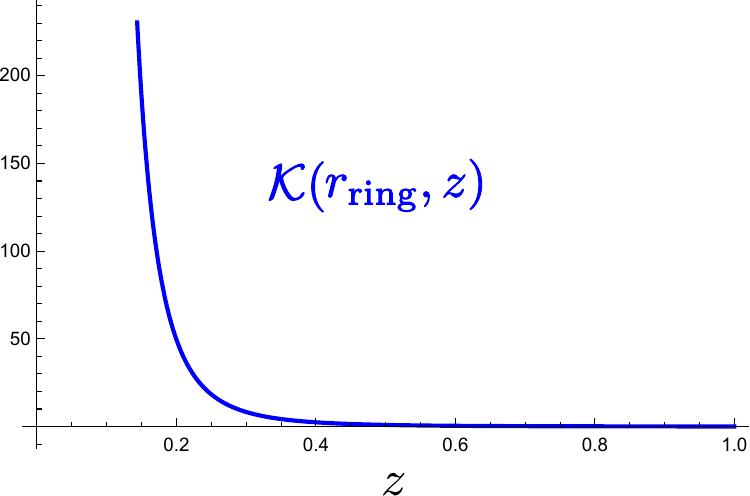}
\caption{On the left panel we represent the Kretschman scalar for a section of the geometry corresponding to a $\varphi = \text{constant}$, for the values of the parameters $M_{\text{ring}} = r_{\text{ring}} = 1$, with the vertical axis representing the $z$-coordinate and the horizontal axis representing the transverse direction. On the right panel we represent the Kretschmann scalar as a function of $z$ for the same choice of parameters along a vertical line that is orthogonal to the ring, to explicitly illustrate how it blows up as $z \rightarrow 0$.}
\label{Fig:RingKretschmann}
\end{center}
\end{figure} 
An additional pathology of these geometries that we have analyzed is the behavior of the Kretschmann scalar. The Kretschmanm scalar diverges at the ring itself as it is illustrated in Fig.~\ref{Fig:RingKretschmann}, leading to a singular geometry (which   is of the naked-singularity type). Thus, even though they are excluded from Israel's analysis by the assumption on the topology of constant redshift surfaces, we see that they cannot represent black holes with regular event horizons. This behavior is generic of any singularity that is not located at the $z$-axis. Thus, we now focus on singularities located exclusively in the $z$-axis.
 
\paragraph*{\textbf{Schwarzschild black hole.}}

The remaining   type of source in the ALP to which we may associate an isolated putative black hole is that of a constant density rod located in the symmetry axis. Taking it to lie between $z = - M$ and $ z = + M$, we have that the functions $U$ and $V$ for this source read:
\begin{align}
    & U_S (r,z) = \frac{1}{2} \log \left( \frac{R_{+} + R_{-} - 2M}{R_{+} + R_{-} + 2M} \right),
    & V_S(r,z) = \frac{1}{2} \log \left( \frac{(R_{+} + R_{-})^2 - 4 M^2 }{4 R_{+} R_{-}} \right). \label{Eq:Schwarzschild}
\end{align}
where we have introduced the following shortcut in the notation
\begin{align*}
    R_{\pm}^2 = r^2 + (z \pm M)^2.
\end{align*}
As is well known, this is precisely the Schwarzschild metric written in Weyl coordinates~\cite{Stephani2003}. The leading order behavior of the functions $U$ and $V$ as $r \rightarrow 0$ for $z \in (-M,M)$ is $\log (r)$. Thus we automatically have the behavior required for a putative black hole, namely,
\begin{align}
    \lim_{\substack{r \to 0 \\ z \in (-M,M)}} U(r,z)= -\infty  .
\end{align}
It has two crucial properties that might appear as surprising from the point of view of the Weyl coordinates, whereas they are well-known features of the Scwharzschild geometry.
On the one hand, the area of the equipotential surfaces, which are ellipsoids in the fiduciary 3-space of the ALP, tends to a finite constant $A= 16 \pi M^2$ when these surfaces approach the rod. On the other hand, the Kretschmann scalar is constant in the equipotential surfaces and also goes to a finite value when the rod is approached. Therefore, the rod itself represents a regular geometrical boundary (i.e. a proper horizon) through which one can extend the geometry. We can see in this way that in this case the source of the ALP problem does not correspond to any kind of matter content from the GR point of view: The Schwarzschild solution can be extended beyond the horizon as a vacuum solution~\cite{HawkingEllis1973}. In the following sections we will analyze  whether it is possible to deform the shape of this black hole by any means and in which cases do such perturbations lead to a singular behavior. 

\subsection{Deforming a black hole from the inside}
\label{Subsec:InsideDefs}
The previous discussion has shown that from the initial list of simple putative black holes only the Schwarzschild geometry represents a proper black hole. By deforming a Schwarzschild black hole from the inside we mean here to eventually perturb the Schwarzschild source term in the ALP problem, maintaining the condition of an empty exterior. Starting from the Schwarzschild solution we can try to see whether there are additional sources apart from the uniform rod leading to a regular horizon. It is clear that we cannot add a distributional point-like structure inside a rod. This will result in a Curzon-like contribution to the geometry that as we saw, leads to irregular horizons. Thus, the only remaining possibility is to modify the linear density of the rod to make it inhomogeneous, i.e. to consider a non-constant function $\lambda(z)$ for the linear density of the rod. 

This situation is carefully analyzed in Appendix B of~\cite{Geroch1982}. Any deformation from the inside of Schwarzschild spacetime needs to be encoded in the presence of inhomogeneities along the initially homogeneous rod, i.e. a function $\lambda(z)$. It translates into a behavior \mbox{$ U = f(z) \log r + o \left( \log r \right)$} as we approach the horizon $r \to 0$ with a fixed $z\in (-M,M)$. Here $f(z)=2\lambda(z)$. This is easy to understand, as when one approaches a segment closely enough, the local linear density dictates the leading behavior of the potential.

When a black hole has a non-regular horizon, a first hint comes from a calculation of its surface gravity
\begin{align}
    \kappa^2 = \lim_{r\rightarrow 0} e^{4U - 2V} \left[ \left( \partial_r U \right)^2 + \left( \partial_z U \right)^2 \right].
\end{align}
It needs to be bounded to have a regular geometry. In fact, we can recall that the surface gravity of a stationary black hole is constant on the horizon due to the first law of black hole dynamics~\cite{Bardeen1973,Racz1995}. Now, boundedness of the surface gravity requires that the density of the rod is homogeneous. Any kind of inhomogeneous behavior leads to a divergent surface gravity. Furthermore, it also implies that the Kretschmann scalar diverges. Let us show this explicitly. Given the leading behavior for $U = f(z) \log r $, as $r \rightarrow 0$ for $z \in (-M,M)$, we have that the radial equation for $V$ becomes:
\begin{align}
    \partial_r V = r \left[ \left(  \frac{f(z)}{r} \right)^2 - \left( f'(z) \log r\right)^2 \right].
\end{align}
Since the second term is of the form $\sim r \log r$, it gives a subdominant contribution as we approach $r \rightarrow 0$ and the leading order behavior of $V$ is similar to that of $U$, namely $V = f^2(z) \log r + \text{subdominant}$. If we plug this ansatz in the surface gravity and the Kretschmann scalar we find:
\begin{align}
     \kappa = & f(z) r^{f(z) - 1} + \text{finite terms}, \label{Eq:Inhomogeneous_Kappa}\\
    \mathcal{K} = &  \frac{12 (f(z)-1)^2 f(z)^2}{r^4} + \nonumber \\
    & \frac{8 f'(z)^2\left[3 f(z)^2 \log ^2(r) -3 f(z) \log ^2(r) +\log ^2(r) +2 \log (r) +2 \right]}{r^2} \nonumber \\
    & + \text{finite terms} \label{Eq:Inhomogeneous_Kretschmann}. 
\end{align}
We see that if $f(z)$ is different from a constant function with a value either $0$ or $1$ we find a singular geometry at $r\to 0$. A possible interpretation is the following. The constant density rod in the ALP represents an auxiliary source and when analyzing the black hole geometry we realize that it does not map to any kind of matter content (the Einstein tensor identically vanishes on the horizon). The inhomogeneous profiles do not correspond to such auxiliary source and do indeed represent a real matter content that arises in a non-vanishing Einstein tensor. Given that it is static (non-violating energy conditions) matter on top of an event horizon, it gives rise to a singular behavior which is what we see in Eqs.~\eqref{Eq:Inhomogeneous_Kappa} and~\eqref{Eq:Inhomogeneous_Kretschmann}. 

Therefore, we conclude that there is no other distributional source in the ALP aside from the homogeneous rod, which gives the Schwarzschild solution, leading to a proper black hole geometry in vacuum (i.e. with a regular black hole). 

\subsection{Deforming a black hole from the outside}
\label{Subsec:OutsideDefs}

Let us consider now more general black holes that are locally in vacuum but do not necessarily extend the vacuum region all the way up to infinity. Geroch and Hartle~\cite{Geroch1982}, showed that every black hole solution with an event horizon whose topology is a two-sphere is such that the function $U(r,z)$ takes the same values at both extremes of the horizon i.e. $U(0,M) = U(0,-M)$ and we have that $U(r,z)$ has the same distributional character as its Schwarzschild counterpart $U_S$ on the event horizon, namely that of a infinitely thin homogeneous rod located on the $z$ axis between $-M$ and $+M$. Equivalently, this means that every local black hole solution can be represented as 
\begin{align}
    U(r,z) = U_S(r,z) + \Delta U (r,z),
\end{align}
with $\Delta U$ any solution to Laplace equation that is analytic and regular, even at the horizon. Such a solution represents a deformed black hole due to an external gravitational field, with the function $V$ being obtained by direct integration. These are all the local black holes in vacuum\footnote{Although they were found previously~\cite{Israel1964,Lawrence1966,Israel1973}, an exhaustive and systematic analysis was presented by Geroch and Hartle~\cite{Geroch1982}, including all the black holes including those with toroidal topology.}. However, as we discussed at the beginning of the section, finding a local solution for $U$ is not the whole story, since we still need to ensure that $V$ has a non-singular behavior. 

The horizon in these coordinates is located in the $z$-axis, between $-M$ and $M$. Hence, the $z$-axis is disconnected and made of two pieces, $  (-\infty,-M) \cup (M,\infty) $. Thus, it is not straightforward that we can ensure that $V(0,z)$ vanishes at both pieces of the axis even if we fix that it vanishes at one of the two connected pieces. The condition $U(0,M) = U(0,-M)$ precisely ensures that $V(0,z)$ vanishes on both connected pieces of the axis if it vanishes on one of them. Indeed, given that we have $U = U_S + \Delta U$, we can also perform an splitting of the form $V = V_S + \Delta V$ and we know that both $U_S$ and $V_S$ obey the vacuum equations. $\Delta V$ should obey the following equation (see Eq.~\eqref{Eq:DzV})
\begin{align}
    \partial_z \Delta V = 2 r \left( \partial_r U_S \partial_z \Delta U + \partial_r \Delta U \partial_z U_S + \partial_r \Delta U \partial_z \Delta U \right). 
\end{align}
Given that $U_S \sim \log r$ as $r \rightarrow 0$, performing an integration parallel to the $z$ axis near $r = 0$, leads to the conclusion that
\begin{align}
    \Delta V(0,M) - \Delta V(0,-M) \sim \Delta U(0,M) - \Delta U(0,-M).
\end{align}
Hence, we can ensure that $V$ vanishes on both pieces of the axis as long as $U(0,M) = U(0,-M)$ (understanding this expression as a limit). This ensures that we can make the function $V$ regular on the whole axis taking it as the integral in Eq.~\eqref{Eq:V_cuadrature}. 

The final step to check is asymptotic flatness, which requires that the system is in equilibrium. If the ALP   of the black hole surrounded by some matter content is in equilibrium, it leads to a solution that is regular and can be extended as a vacuum solution all the way up to infinity, with the matter located only on compact regions.

\begin{figure}[H]
\begin{center}
\includegraphics[width=0.75 \textwidth]{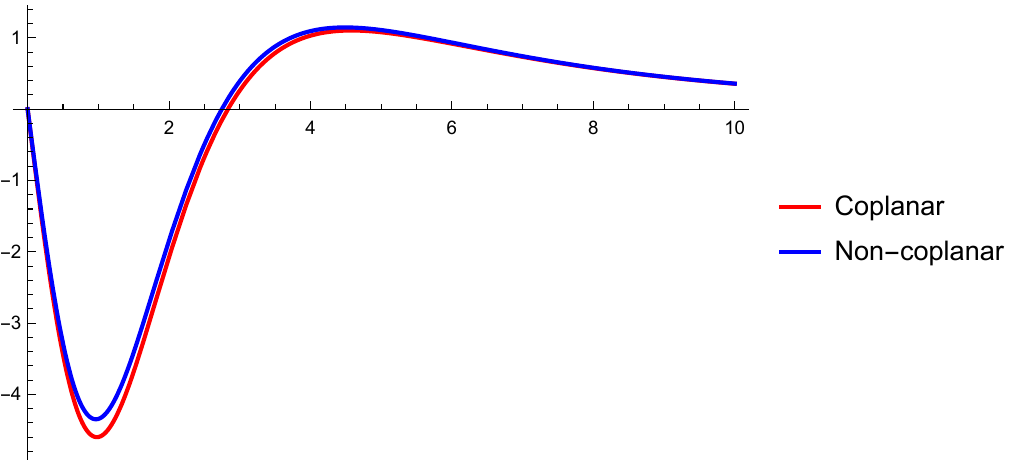}
\caption{Here we represent the integrand of \eqref{Eq:Equilibrium_Condition} a function of $r$ for some specific choices of the parameters, namely we are depicting the value for $z = 3$ and we are taking the radius and mass of the ring to be $M_{\text{ring}} = 1, r_{\text{ring}}= 1$ and the black hole mass $M= 1 $ . The red curve corresponds to the integrand when the ring and the black hole are on the same plane, whereas the blue one corresponds to them lying on different planes, namely five units displaced in the z-axis. A priori, there is no apparent symmetry reason to expect a cancellation of the integral.}
\label{Fig:Saturns_Integrands}
\end{center}
\end{figure}

An illustrative example that we can explicitly construct is precisely a black hole deformed by the presence of an external ring of matter. For details about how the functions $U$ and $V$ for the so-called Weyl ring are constructed, see Appendix~\ref{App:Rings}:
\begin{align}
    U = U_S + U_{\text{ring}}. 
\end{align}
The ring can have a thickness, but for computational purposes, it is convenient to consider an infinitely thin ring. In this case, we have to ignore the singular behavior at the ring itself, as it is an artifact of the approximation. If the ring is located at a plane that intersects the black hole symmetrically, the resulting geometry is perfectly regular and the integral~\eqref{Eq:Equilibrium_Condition} vanishes since the configuration is in equilibrium in the $z$ direction, i.e., $\partial_zU=0$. The stability of the ring itself to its own shrinking is provided by internal pressures that can be directly obtained from the Einstein equations. However, if the ring is located on a different plane,   the equilibrium condition \eqref{Eq:Equilibrium_Condition} is not satisfied and hence regularity at the axis and asymptotic flatness cannot hold simultaneously.\footnote{Although now the ALP is associated with a homogeneous rod at the axis and not to pure vacuum, the same argument applies if we slightly deform the contour $\gamma_z$ in Fig.~\ref{Fig:Sketch2} to avoid the distributional matter content representing the horizon.}
\begin{figure}[H]
\begin{center}
\includegraphics[width=0.25 \textwidth]{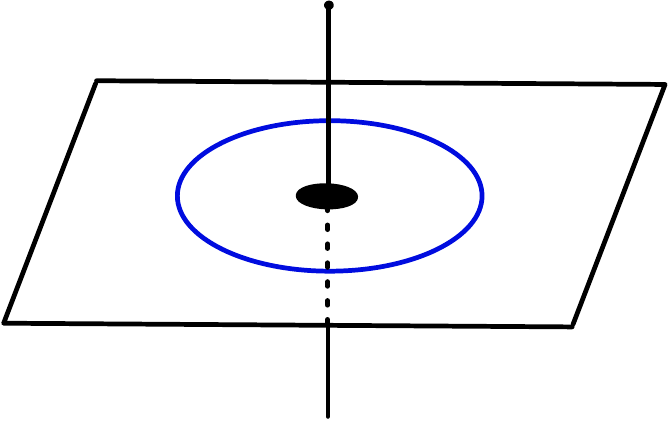}
\includegraphics[width=0.25 \textwidth]{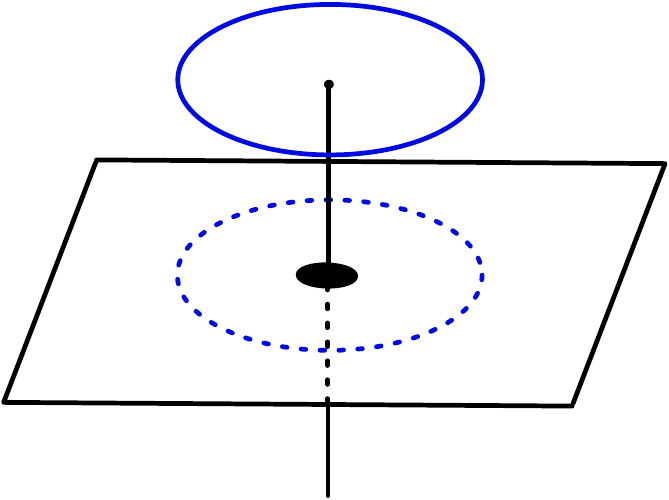}
\caption{The shadowed black region represents the black hole, and the blue ring can be located either in the same plane as the black hole, or in a plane that is not the one of the black hole. In the former case we find that the integral vanishes, whereas in the latter case we find a finite result.}
\label{Fig:Saturns}
\end{center}
\end{figure} 
For details about how the  functions $U_{\text{ring}}$ and $V_{\text{ring}}$ for the so-called Weyl ring are constructed, see Appendix~\ref{App:Rings}. We have not been able to perform the integral~\eqref{Eq:Equilibrium_Condition} analytically in either case and it would be quite unlikely to conclude that the integral vanishes simply by inspecting the integrand if we did not have the equilibrium interpretation, see Fig.~\ref{Fig:Saturns_Integrands}. 
More in general, we can take any kind of matter content in equilibrium surrounding the black hole and consider a deformation which is given directly by solving the ALP with the suitable boundary conditions that correspond to the matter content decided. In fact, the boundary conditions associated with the matter content can be replaced with an effective density $\rho_{\text{eff}} (\boldsymbol{x})$ that we can integrate with the Green function $G(\boldsymbol{x},\boldsymbol{x'}) = 1/4 \pi \abs{\boldsymbol{x} - \boldsymbol{x'}}$ to find the contribution to the $U$ function given by the matter content:
\begin{align}
    \Delta U (\boldsymbol{x})= \int d^3 x' \frac{\rho_{\text{eff}} ( \boldsymbol{x'})}{\abs{\boldsymbol{x}-\boldsymbol{x'}}}.
\end{align}
We find that as long as there is no density in the horizon, we are avoiding the coincidence limit in the Green function and no singular behavior arises.

\section{Black holes in astrophysical environments}
\label{Sec:NoHair_External}

After an extensive discussion of the potential deformations of black holes in the restricted case of static and axisymmetric configurations, we are now in a position to formulate our no-hair result for black holes in astrophysical environments. 

To the best of our knowledge, the only previous results in the literature that go in this direction are those of G\"urlebeck~\cite{Gurlebeck2012,Gurlebeck2015}. G\"urlebeck showed that the Weyl multipoles of spacetime can be expressed as a sum of different contributions in the form of integrals that are localized in compact regions around coordinate singularities (e.g. horizons or defects) and regions containing some matter content. Furthermore, G\"urlebeck showed that the contribution from black-hole-like singularities is always the same as the one from Schwarzschild isolated black holes. Here we are going to present the no-hair theorem in a different way which constrains the form of the metric itself.

\noindent
\textbf{Theorem:} Given a regular (in the sense precised below) gravitational environment, there exists only one static, axisymmetric and asymptotically flat geometry containing a black hole which is non-singular. Moreover, the horizon of the black hole will depart from spherical symmetry (the shape when the environment is trivial) in a unique and specific manner that is completely dictated by the environment.

As a first step, let us characterize what we mean by an external gravitational environment in the restricted setup of staticity and axisymmetry that we are considering. 

\begin{figure}
\begin{center}
\includegraphics[width=0.35 \textwidth]{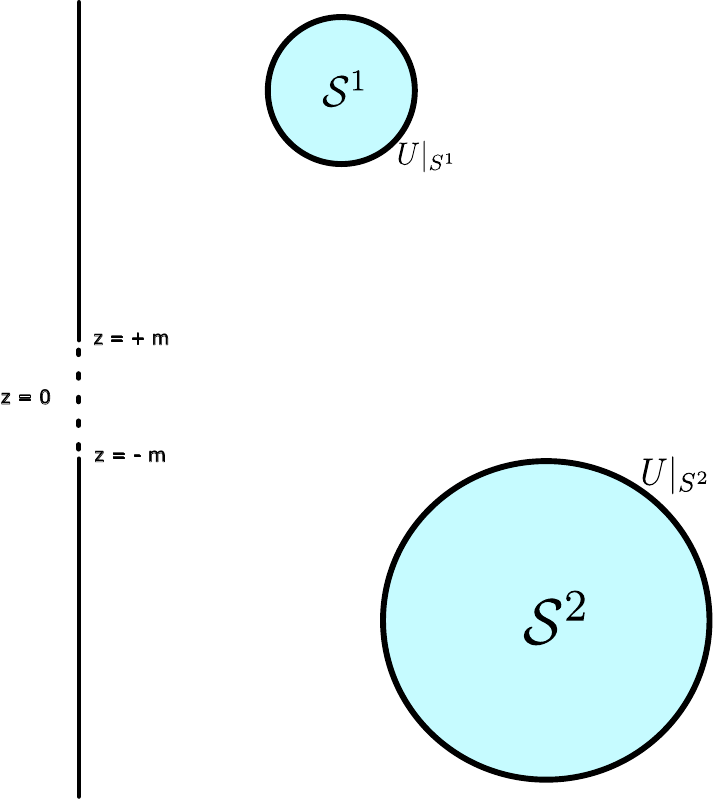}
\caption{Schematic representation of the setup. The $z$-axis is depicted as a black solid line, with a dashed segment between $z=-M$ and $z=M$ where the event horizon is located. Finally, we have illustrated some compact surfaces enclosing matter content (they would correspond to topological torus once we consider the $\varphi$ direction too) and we replace the potential matter content inside with a boundary condition at the surface of the object.}
\label{Fig:BH_screens}
\end{center}
\end{figure}

Consider a homogeneous rod source for the ALP located at $r=0$ between $z = M$ and $z = -M$ in the coordinate system that we are using. Furthermore, consider a set of compact and $\mathcal{C}^{\infty}$ toroidal two-surfaces\footnote{While it may be possible to consider less regular surfaces, we assume them to be arbitrarily smooth for simplicity in our analysis.} $S^i$ enclosing the external regions containing matter, see Fig.~\ref{Fig:BH_screens} for a pictorial representation. Inside these surfaces the vacuum equations are not obeyed and a careful analysis of the matter content would be required to determine the metric inside. However, to analyze their influence outside, we can replace the matter content by the value of the function $U$ induced on those surfaces. We restrict ourselves here to situations in which the functions $U$ at these surface are at least $\mathcal{C}^2$.\footnote{Although it must be possible to relax this condition, this is enough for our purposes.} Notice that we are not only demanding that the $U$ restricted to the surface is $\mathcal{C}^2$ but also across the surface, which automatically excludes the presence of any distributional matter content, i.e. thin shells. Then, the values of $U$ at the different toroidal surfaces act as Dirichlet boundary conditions.\footnote{This is similar to the analysis of systems like the electromagnetic field between mirrors, where a perfectly reflecting boundary condition is imposed as a way to model in a simple way the interaction between the field and the matter content in the boundary. In this way, we can focus only on the determination of the field from its equations of motion supplemented with this boundary condition.}

As we have shown in the previous section, the central black hole can only be represented by a homogeneous rod source with a single free parameter: its length or mass parameter $M$. Knowing this internal source, the value of the function $U$ at the toroidal two-surfaces, and the asymptotic condition that $U \to 0$ at infinity to ensure asymptotic flatness, we can solve the ALP equation for all vacuum regions in spacetime. In fact, we have a well-posed elliptic problem with Dirichlet boundary conditions, i.e., we know the harmonic function $U$ at the horizon, at the surfaces $S^i$, and we know that it goes to zero at infinity as a polynomial in $1/r$. Hence, the solution to the Laplace equation exists and is unique outside the surfaces by virtue of the classic theorems~\cite{Gilbarg2001}.

The final ingredient that we need to ask for is consistency, i.e., that the equilibrium condition~\eqref{Eq:Equilibrium_Condition} is obeyed everywhere. This is required so that the system does not display any kind of singular (cosmic-string-like) behavior, and we are really representing a black hole configuration in a gravitational environment. The specific form of the function $\Delta U=U-U_S$ determines the specific departure from spherical symmetry of the distorted black hole. 

Putting everything together, we have shown that:
\begin{enumerate}
    \item Given a gravitational environment, the horizon of a black hole departs from spherical symmetry in a unique and specific manner dictated by the environment. Any deviation from this natural shape will result in an irregularity of the geometry.
    \item Any spheroidal shape for the horizon is in principle possible; one only needs to find which would be the necessary gravitational environment.
    \item Deformations of the horizon generated by external matter content are mild in terms of curvature, in the sense that they give rise to bounded curvature contributions at the horizon.
    \item In any case, for a solution with black holes and external matter to be regular everywhere, we need the equilibrium condition \eqref{Eq:Equilibrium_Condition} in the ALP. 
\end{enumerate}

\section{Almost-no-hair results for horizonless ultracompact objects}
\label{Sec:NoHair_ECOs}

For static and axisymmetric black holes we have shown that we can find a uniqueness result under complete generality, i.e. without some of the conditions in Israel theorem, such as the constraint on the topology of the constant redshift surfaces and the finiteness of the area of the minimum redshift surface, which appear necessary in the general static case. Now, let us consider the scenario in which, instead of black holes, we are dealing with horizonless ultracompact objects. By this, we refer to material bodies whose first equipotential surface outside matter has an extremely high but finite redshift, characterized by a parameter $e^{2U_{\rm surf}} = \epsilon$, where $0 < \epsilon \ll 1$.

Given that no-hair theorems require the existence of a horizon one could think that, if no horizon is present, any ultracompact object would be possible, for instance compact but finite Curzon objects, thin rings, and spheroidal but not spherically-symmetric ultracompact objects. However, a new form of no-hair results can be obtained by requiring that the Kretschmann scalar does not acquire trans-Planckian values~\cite{Barcelo2019}, i.e. ${\cal K}< {\cal K}_P=1/\ell_P^4$ with $\ell_P$ representing the Planck length. More generally, this condition aligns with the limiting curvature hypothesis, which postulates that all curvature invariants should remain bounded~\cite{Markov1982,Markov1987,Frolov1988,Frolov2021,Frolov2022}. This curvature constraint effectively replaces the standard condition of a regular horizon in traditional no-hair theorems. As we will demonstrate, this constraint imposes stringent limitations on the permissible geometries, particularly in terms of the degree of distortion possible from spherical symmetry. The underlying physical motivation is that  GR  itself becomes unreliable when curvatures reach such extreme values. Indeed, this curvature constraint excludes not only non-spherical geometries but also spherically symmetric ultracompact configurations at the Planck scale from the set of plausible spacetime structures. This contrasts with the standard Israel theorem for black holes, which remains indifferent to the exact curvature values, as long as it remains bounded. In general, the GR dynamics is such that it tends to avoid configurations with extremely high curvatures except in cases where gravity becomes so strong that singularities inevitably form. Hence, it is reasonable to expect that any viable theory extending GR would naturally favor configurations with lower curvatures if such alternatives are accessible, thus avoiding configurations with extremely large curvatures. 

As we saw before, in the static and axisymmetric case, there are only three possible sources in the ALP that can lead to arbitrarily high redshifts. Let us now analyze these three cases separately.

\paragraph*{\textbf{Curvature-induced bounds for ultracompact lenticular objects.}}

When dealing with metrics close to the Curzon family we can think that these metrics describe the external region of an object with finite extent instead of a point-like region in Weyl coordinates. Outside a very small radius $R=R_0$ encompassing the object, the geometry can be defined by a specific combination of Weyl multipolar contributions. As described in Sec.~\ref{sec:curzon} above, the Kretschmann scalar always grows unboundedly in some concrete directions depending on the specific multipolar expansion of the configuration.

Let us start analyzing the simplest monopole contribution, namely $\ell=0$ in Eq.~\eqref{Eq:U_Multipoles}. Given a fixed value of $\epsilon$ representing the minimum redshift attained by the configuration, we have a relation between $M^{(0)}$, $R_0$, and $\epsilon$ of the form 
\begin{align}
    \epsilon = \exp \left( -\frac{2M^{(0)}}{R_0} \right).
\end{align}
The ratio $M^{(0)}/R_0$ must be large in order to have $\epsilon \ll 1$. We can express $M^{(0)}$ in terms of  $\epsilon$ and $R_0$ as:
\begin{align}
    M^{(0)}=\frac{1}{2} R_0 \abs{\log \epsilon}.
    \label{Eq:LogSeparation}
\end{align}
Now, the maximum value of the Kretschmann scalar in the Weyl sphere of radius $R_0$ is
\begin{align}
    \mathcal{K}_{\rm max} \sim \frac{1}{{\cal R}_0^4} \exp\left( \frac{ 2 [M^{(0)}]^2 }{R_0^2} \right),
\end{align}
where we have neglected subdominant terms for small $R_0$. The limitation on the maximum value of the Kretschmann amounts to an inequality of the form  
\begin{align}
\frac{{\cal R}_0}{\ell_P} \gtrsim \exp\left( \frac{[M^{(0)}]^2 }{2R_0^2} \right) =
\exp\left( \frac{1}{8} |\log\epsilon|^2 \right).
\end{align}
For very small $\epsilon$ we find that 
\begin{align}
\frac{R_0}{\ell_P} \gtrsim \frac{1}{\sqrt{2}} \abs{\log \epsilon}^{3/2} \exp\left( \frac{1}{8} |\log\epsilon|^2 \right),
\label{Eq:ParametricalSeparation}
\end{align}
after substituting the specific function $\mathcal{R}_0 \left( R, \pi/2; M^{0}\right)$ from Eq.~\eqref{Eq:MathcalR}. We are taking $\theta = \pi/2$ which is the direction in which the Kretschmann grows faster as we approach $R \rightarrow 0$.
In conclusion, a monopolar Curzon-like object with a minimum redshift surface (characterized by a small $\epsilon$) must be extremely large (i.e., $R_0 \gg \ell_P$), with an even larger $M^{(0)}$, to avoid high curvature values. This implies the following hierarchy of scales: $M^{(0)} > R_0 \gg \ell_P$. While the separation between $M^{(0)}$ and $R_0$ depends logarithmically on $\epsilon$ and therefore does not need to be exceptionally large [see Eq.~\eqref{Eq:LogSeparation}], the separation between $R_0$ and $\ell_P$ must [see Eq.~\eqref{Eq:ParametricalSeparation}].

\begin{figure}[H]
\begin{center}
\includegraphics[width=1 \textwidth]{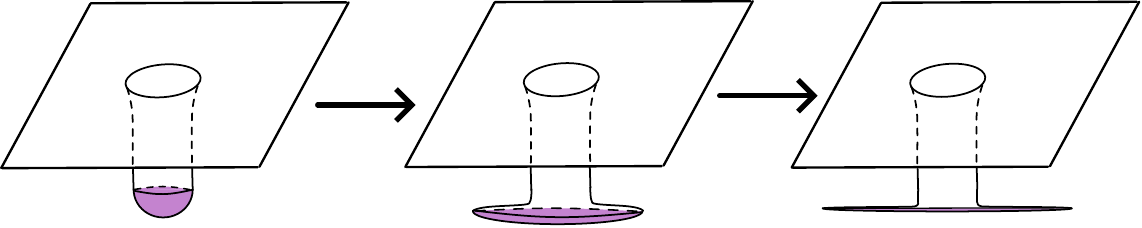}
\caption{Pictorial representation of gravitational collapse into a Curzon-type metric. The colored region represents the material content. Rather than generating an increasingly deep gravitational potential well, the collapse compresses the matter into a plane and stretches it along that plane.}
\label{Fig:Lenticular_Collapse}
\end{center}
\end{figure}

As we have already mentioned, an ultracompact monopolar Curzon object is actually not spherically symmetric even if the redshift function $U$ is spherically symmetric. In fact, it could be seen as representing the collapse of a cloud of matter tending towards a very big and thin lenticular object (see the illustrative picture in Fig.~\ref{Fig:Lenticular_Collapse}). It corresponds to an anisotropic collapse in which matter is squeezed in one direction while it spreads over in the other two. However, our analysis shows that, given a mass for the lenticular object, it cannot be too flattened to avoid the presence of trans-Planckian curvatures at the lentil's rim.
An equivalent discussion applies to any other ultracompact object based on a different multipolar combination. The difference now is that equipotential surfaces are no longer precise spheres in Weyl coordinates, resulting in objects that can have more complex shapes.
\paragraph*{\textbf{Curvature-induced bounds for ultracompact rings}}
We have found a quite convoluted expression for the Kretschmann scalar using the software \textit{Mathematica} with the help of the xAct package~\cite{xAct}. We have attached a Mathematica notebook containing the explicit expression which is not very illuminating. Thus, we have decided to perform an analysis based on purely dimensional grounds and inspecting the leading orders of the divergences in the potential and Kretschmann scalar. Qualitatively speaking the analysis is very similar to the previous case.

Close to the ring, the equipotential surfaces are tori. For a given value of the mass $M$, the minimum value of the redshift $\epsilon$ is acquired for a torus of given radius $d$. In reality, there is a dependence on the direction from which one approaches the center of the ring, but we ignore this dependence here, considering $d$ as an average distance to the center of the regular ring. We can understand this radius as the thickness of the ultracompact ring under consideration. The leading order of the redshift function imposes a relation between the mass of the ring and such radius:
\begin{align}
\epsilon \sim \exp(-M/d) \quad\Longrightarrow\quad M \sim d   \abs{ \log \epsilon}.
\end{align}
We can think that the mass of the ring is fixed by $\epsilon$ and $d$. Now, the bound on the curvature implies
\begin{align}
{\cal K} \sim e^{N(M/d)^n}{\cal R}^{-4} < \frac{1}{\ell_P^4},
\end{align}
where we have substituted the Kretschmann by its leading order expansion expected on dimensional grounds. The length scale ${\cal R}$ has a very convoluted form. 
Inserting the previous expression for $M$ as a function of $d$ and $\epsilon$ we will obtain a condition of the form
\begin{align}
d > \ell_P F (\abs{\log \epsilon}),
\end{align}
with $F$ a growing function of its argument $\abs{\log \epsilon}$ which tends to infinity as $\epsilon \to 0$. At the end, we would find again a separation of scales $M > d \gg \ell_P$. 

\begin{figure} 
\begin{center}
\includegraphics[width=0.80 \textwidth]{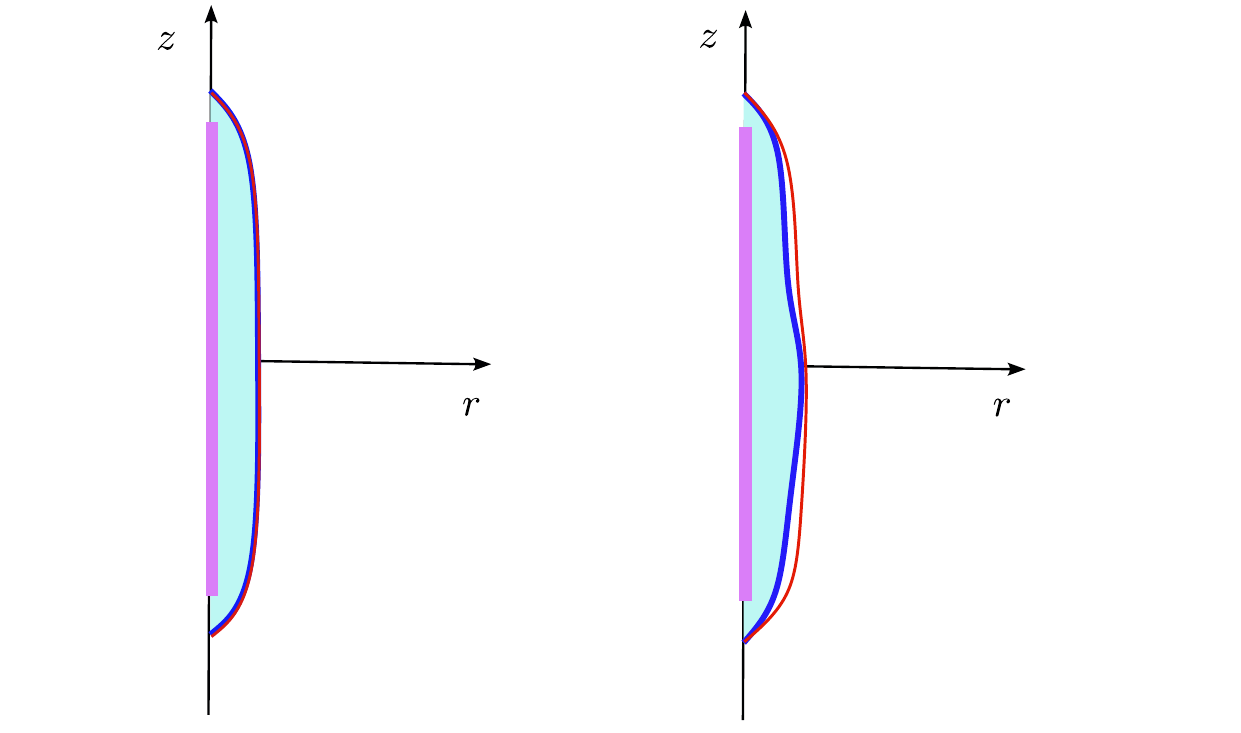}
\caption{The figure, presented in Weyl coordinates [using only the $(r,z)$ plane], illustrates the contrast between a spherically symmetric ultracompact star (left) and a distorted configuration (right). For reference, we have included the homogeneous rod source representing the pure Schwarzschild solution. A spherically symmetric ultracompact star is one whose surface (blue line) precisely aligns with an equipotential surface of the Schwarzschild solution close to the rod (coinciding red line). By contrast, a non-spherically symmetric object exhibits equipotential surfaces that are distorted relative to the Schwarzschild case, and its surface does not necessarily coincide with any equipotential surface. In the right panel, the red line represents the innermost equipotential surface external to the material body. }
\label{Fig:EUCSpheroids}
\end{center}
\end{figure}

\paragraph*{\textbf{Curvature-induced bounds for ultracompact spheroids}}
As a final case, let us examine the most relevant situation from the astrophysical point of view. It corresponds to ultracompact stellar-like objects that can mimic most properties of general relativistic black holes due to having a deep gravitational well, which makes them almost observationally indistinguishable from black holes. Consider first an ultracompact but spherically symmetric object (see the left-hand side of Fig.~ \ref{Fig:EUCSpheroids}). Its surface, which is a constant redshift surface $\epsilon$, represents an ellipsoid in Weyl coordinates. However, we know that this is an artifact of using Weyl coordinates and that, from the point of view of the spacetime geometry, it constitutes a perfectly spherically symmetric surface. To prevent trans-Planckian curvatures in this spherically symmetric setup, it suffices for the object to be macroscopic, meaning $M \gg M_P$ (which, in our units $G=c=1$, corresponds to $M \gg l_P$). The compactness parameter $\epsilon$ becomes irrelevant when we consider only spherically symmetric configurations, as these can be made arbitrarily compact without incurring excessively high curvature values.

Now, let us consider deviations from spherical symmetry. We define an object as ultracompact if, outside its matter content, the innermost equipotential surface takes on a very small value, $\epsilon\ll 1$. This surface is distorted compared to the equipotential surfaces of spherically symmetric objects (see the right panel in Fig.~\ref{Fig:EUCSpheroids}). Furthermore, in general, the object's surface does not necessarily coincide with an equipotential surface. In GR, and in the static case, deviations from sphericity are effectively described using the Geroch multipoles~\cite{Geroch1970} (in stationary situations these multipoles generalize to the so-called Geroch-Hansen multipoles~\cite{Hansen1974}). The set of multipoles consistent with axisymmetry form a series characterized by a single number $\ell$. The strength of each multipole is characterized by a real number $M_{\rm G}^{(\ell)}$, with the monopolar contribution representing the mass of the object, $M_{\rm G}^{(0)}=M$.

It is interesting to point out a subtlety in the relation between Weyl and Geroch multipoles. Far from a matter source, an expansion in Weyl multipoles can always be mapped to an expansion in terms of Geroch multipoles. In particular, a Geroch multipole of order $\ell$ can be decomposed as a sum of Weyl multipoles [defined in Eq.~\eqref{Eq:U_Multipoles}]:
\begin{align}
    M_{\rm G}^{ (\ell)} = \sum_{n=1}^{\ell} \sum_{k_{1}...k_{n}} c_{k_{1}...k_{n}} M^{(k_{1})} \times \cdots \times M^{(k_{n})}; \qquad \sum_{i=1}^{n} k_{i} = \ell,
\end{align}
where the second equality is a constraint over the sum $\sum_{k_{1}...k_{n}}$, and $c_{k_{1}...k_{n}}$ are coefficients, for which there is not an explicit closed expression, but which can be found at any finite order through the algorithm presented in~\cite{Fodor1989}. However, the entire field structure associated with a Geroch multipole, i.e., the full exterior region of the Schwarzschild metric and not only the region far from the event horizon, cannot be described everywhere outside the event horizon as a Weyl multipolar expansion. Specifically, the function $U_S$ cannot be expressed as a series of Weyl multipoles centered at a single point. In fact, this shows that the Weyl multipolar expansion is not well-suited for describing spherically symmetric configurations

\begin{figure} 
\begin{center}
\includegraphics[width=0.80 \textwidth]{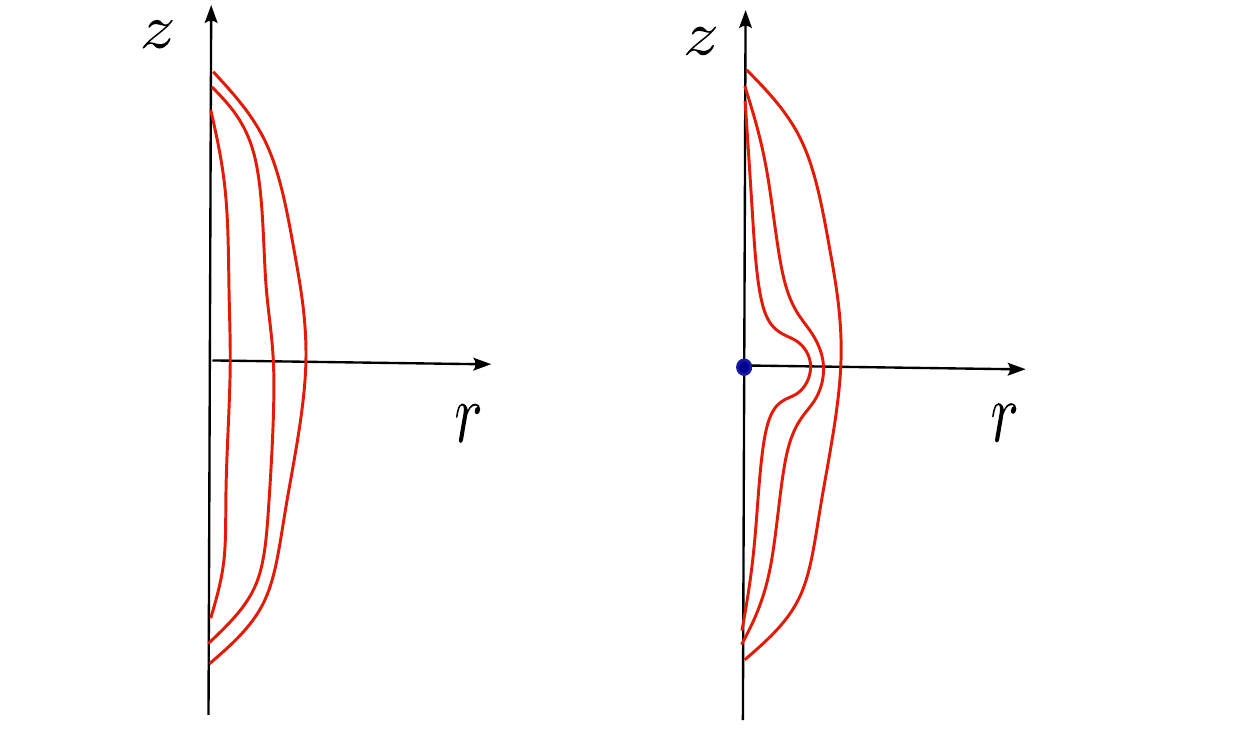}
\caption{The red lines in the figure represent in Weyl coordinates the innermost equipotential surfaces outside the body for different ultracompact objects,. The closer the surface to the axis of symmetry the bigger the compactness of the object.
The figure on the left represents a series of more and more compact objects tending to a limit described by an inhomogeneous rod, with a linear density function $\lambda(z)$. The figure on the right represents instead a series of ultracompact configurations which, apart from an inhomogeneous rod, unveil a singular Weyl monopolar contribution. Essentially, these are the two types of tendencies one could find when analyzing the geometry of ultracompact objects with higher and higher compactness.}
\label{Fig:Sch-vs-Cur}
\end{center}
\end{figure}

It is evident that different ultracompact material distributions can produce the same external multipolar expansion. In other words, a given gravitational field generated by a matter distribution (noting that there is no horizon, ensuring that it is causally connected with the exterior region) can arise from infinitely many different internal matter configurations. However, for isolated objects and in the singular $\epsilon \to 0$ limit, the distortions (which actually become infinite on the $\epsilon = 0$ surface) from the pure Schwarzschild form can only come about essentially in two essential forms. Ultracompact objects either tend to an inhomogeneous rod (see the pictorial representation on the left of Fig.~\ref{Fig:Sch-vs-Cur}) or they acquire some additional Weyl multipolar contributions localized at isolated points of the rod (see the illustrative drawing on the right of  Fig.~\ref{Fig:Sch-vs-Cur}). It all depends on whether making $\epsilon$ smaller unveils a $f(z) \log(r)$ behavior of the redshift function or the $1/r^{\ell+1}$ behavior associated with a Weyl multipole contribution. In the latter case, the geometry is dominated by the Weyl multipoles as $\epsilon \to 0$ so it would be more appropriate to talk about distorted Curzon objects than to talk about distorted Schwarzschild black holes. The curvature condition will impose on these objects the constraints that we already discussed for ultracompact Curzon-like objects. Hence, we focus now on ultracompact objects whose leading order behavior for $U$ is $f(z)\log r$ as $r \rightarrow 0$ for $z \in (-M,M)$.
As we are going to see, $f(z)$ needs to become closer to $1$ everywhere in $z \in (-M,M)$ as we make $\epsilon$ smaller in order to avoid the generation of trans-Planckian curvatures. If we think of a distortion with respect to an ultracompact spherically-symmetric object of mass $M$, we can impose additional constraints on $\lambda(z)$. In particular, we can impose that: i)~the total mass of the system is fixed, and ii) that its center of mass is located at the origin of coordinates $(r,z)=(0,0)$. This is equivalent to:
\begin{align}
\int_{-M}^{+M} dz\,[f(z)-1]=0;~~~~~
\int_{-M}^{+M} dz\,[f(z)-1] z =0.
    \label{Eq:lambdaexpansion}
\end{align}
For an ultracompact stellar-like object we have, at leading order in the limit $\epsilon \to 0$, 
\begin{align}
  \epsilon \sim \exp[2 f(z) \log (r_e/M)];~~~~~~\epsilon \sim \left( \frac{r_e}{M}\right)^{2f(z)}.
  \label{Eq:redshif-spheroids}
\end{align}
Here, $r_e=r_e(\epsilon,z)$ is the location of the innermost equipotential surface outside the object at a given $z$.
We see again that for $\epsilon \ll 1$ we need a large mass-over-size ratio. Now, the leading order behavior of the Kretschmann scalar for this kind of configurations is 
\begin{align}
   \mathcal{K} \sim  \frac{12[f(z)-1]^2 f(z)^2}{r_e^4}
   \label{Eq:Inhomogeneous_Kretschmann_leading}. 
\end{align}
Using Eq.~(\ref{Eq:redshif-spheroids}), the curvature bound leads to the condition
\begin{align}
   \frac{1}{\ell_P^4} > \frac{12[f(z)-1]^2 f(z)^2}{r_e^4}, 
\end{align}
or, equivalently,
\begin{align}
   \frac{r_e}{\ell_P} > \left\{12[f(z)-1]^2 f(z)^2\right\}^{1/4}.
   \label{Eq:size-condition}
\end{align}
For $f(z)=1$ the value of $r_e$ is not constrained, which is precisely the case in which the configuration is spherically symmetric. Hence, a strictly spherically symmetric object can be as compact as desired without incurring in Planckian curvatures, as long as it remains macroscopic ($M \gg \ell_P$). On the other hand, when $f(z) \neq 1$, the bigger the departure of $f(z)$ from unity, the larger would have to be the ultracompact object if one does not want to generate trans-Planckian curvatures. Notice that Eq.~\eqref{Eq:lambdaexpansion}, which fixes the monopole, automatically excludes the case in which $f(z) = 0$ everywhere. Actually, $f(z)$ can become zero at some points, although for instance, if it is a simple zero, the subleading term in Eq.~\eqref{Eq:Inhomogeneous_Kretschmann} would still impose constraints on $f'(z)$. This remains true for some values of $z$ even in the case in which $f(z) = 0$ is zero on an interval $I \subset (-M,M)$, which physically would represent two disconnected objects separated by a vacuum region in between.

There is an alternative way to state the previous result. Let us consider that both the mass of the object and $\epsilon$ are fixed. Then, due to Eq.~\eqref{Eq:redshif-spheroids} the value of $r_e$ is also fixed. Now Eq.~\eqref{Eq:size-condition} must be read from right to left:  
\begin{align}
\left\{12[f(z)-1]^2 f(z)^2\right\}^{1/4} < \frac{r_e}{\ell_P}= \frac{M}{\ell_P} \epsilon^{1/2f}
   \label{Eq:multipole-condition}. 
\end{align}
That is, the non-sphericities of the ultracompact object must be smaller than the quantity $r_e/\ell_P$, which in turn is proportional to (some positive power of) $\epsilon$. Then, as the object becomes more and more compact, its degree of non-sphericity has to die off or trans-Planckian curvatures will be generated.

There needs to exist a one-to-one correspondence between the functions $f(z)$ and the Geroch multipoles $M_{\rm G}^{ (\ell)}$, with $\ell > 1$, of the ultracompact object. The specific functional relation can be however quite convoluted and we do not aim to present it here. We will just sketch for the sake of completeness how it could be established, and leave it for future work. 

First of all, we notice that the Weyl form of the Schwarzschild metric in Eq.~\eqref{Eq:Schwarzschild} can be transformed into the standard Schwarzschild coordinates $(u,\theta)$ by performing the transformation
\begin{align}
    r=\sqrt{u(u-2M)} \sin \theta;~~~~~ z=(u-M)\cos \theta. 
\end{align}
Applying this transformation blindly to a generic Weyl metric~\eqref{Eq:Line-Element-v2} we obtain
\begin{align}
    ds^2 = & - e^{2 U} dt^2  \nonumber \\
    & + e^{-2 U} \bigg[ e^{2V} \left( (1 + \frac{M^2 \sin^2 \theta }{u^2-2Mu}) du^2 + \left( u^2-2Mu + M^2 \sin^2 \theta \right) d\theta^2 \right) 
    \nonumber \\
    &  + u(u-2M) \sin^2 \theta d \varphi^2 \bigg],
    \label{SchwCoor}
\end{align}
where the functions $U$ and $V$ need to be expressed in terms of $u$ and $\theta$. The potential $U$ associated with an inhomogeneous rod of density $\lambda(z)$ is given by
\begin{align}
    U(r,z) = - \int_{-M}^{M} dz' \frac{\lambda(z')}{\sqrt{r^2 + (z'-z)^2}}.
\end{align}
Let us recall that, for a fixed $z \in (-M,M)$ in the limit $r \to 0$, we have 
\begin{align}
    U(r,z) \simeq  2 \lambda(z) \ln r = f(z) \ln r.
    \label{Eq:Uleading}
\end{align}
As discussed at the beginning of the article, from $U$ one can directly obtain $V$ through quadratures (although in general they cannot be written in closed form) and then plug them into the line element in Eq.~\eqref{SchwCoor}. Because of the previous conditions on the center of mass, we can be sure that 
the Schwarzschild coordinates that we are using are Thorne's ``Asymptotically Cartesian and Mass-Centered" (ACMC) coordinates~\cite{Thorne1980}. Then, from the form of the metric one could extract Thorne's multipoles, and in turn  Geroch's multipoles~\cite{Gursel1983}. Deviations from the Schwarzschild monopolar contribution that preserve the center of mass are described by the $\ell >1$ Geroch multipole decomposition. This establishes the connection between the inhomogeneity function $\lambda(z)$ and Geroch multipoles. It is then clear that Geroch multipoles have to die off as compactness becomes very small to avoid the generation of trans-Planckian curvatures, as the function $f (z)$ tends to $1$.

Let us now mention and briefly compare with our findings two previous results in the same direction. In the analysis presented in~\cite{Barcelo2019}, the static case is studied without additional assumptions. There it is shown that, under the same conditions imposed here, the Geroch multipole structure of the configuration parametrically approaches that of the Schwarzschild solution as the redshift $\epsilon$ tends to zero. 
The paper was restricted to a perturbative analysis of the possible deformations around the Schwarzschild solution that have a smooth limit to it in the $\epsilon \to 0$ limit. This leaves the set of nonperturbative solutions that do not have such a smooth limit outside the analysis, which in the case under consideration here are easily analyzed since they constitute the whole Curzon sector. In the static and axisymmetric case considered here, we have been able to completely constrain the structure of different ultracompact metrics including deformations of spherically-symmetric ultracompact spheroids, which are the ones analyzed in this reference. 

Something similar happens with the analysis in~\cite{Raposo2019}. They study linear perturbations on top of a patch of the Schwarzschild metric. Specifically, they consider the Schwarzschild solution with mass $M$ from a surface of proper radius slightly larger than the would-be event horizon, i.e., the region between $r = 2M (1 + \epsilon)$ and $r \to \infty$. There it is shown that the boundedness of curvature invariants leads to a parametric suppression of the Geroch multipole deviations from the Schwarzschild solution. They find that all Geroch-Hansen mass multipoles vanish polynomially in $\epsilon$ if they are ``non-spin induced", meaning that they vanish as the first current multipole of the leading-order perturbation goes to zero. Since we restrict ourselves to the static case, these ``non-spin induced" multipoles correspond to the same notion we have analyzed in here. Thus, in the intersecting domain of validity of these analyses, their conclusions and ours are equivalent.

To conclude this section, it is important to highlight that, for an ultracompact object with a small but fixed $\epsilon,$ it is not strictly possible to determine whether the distortions from strict sphericity originate from within the object or from external influences, as both sources would lead to finite distortions. To distinguish between these possibilities, one would need to probe the external gravitational background and assess whether the object exhibits distortions beyond those naturally caused by external gravitational effects. What we have demonstrated here is that any distortion originating from within the object must be very small—the smaller the more compact the body is.

\section{Conclusions and further remarks}
\label{Sec:Conclusions}

In this work we have analyzed in an exhaustive manner the uniqueness characteristics of black holes under the assumptions of staticity, axisymmetry, and asymptotic flatness. One the one hand, we have presented results scattered in the literature in a systematic and novel manner. On the other hand, we have extended these results to black holes immersed in an external gravitational environment and to horizonless ultracompact objects. 

The axisymmetric case allows one to show that the only vacuum and asymptotically flat black hole solution with a regular horizon is Schwarzschild spacetime. When compared with Israel theorem~\cite{Israel1967} for static geometries, we clearly see that the axisymmetry restriction allows to get rid of two premises of the theorem: the spheroidal topology of the equipotential surfaces condition, and the finite horizon area condition. 

Then, we have discussed the difference between deforming a black hole from the inside or from the outside. We have clearly seen that, while it is not possible to deform a black hole horizon from the inside without making it singular, it can indeed be deformed from the outside. These latter deformations are caused by the gravitational fields of external environments.
To the best of our knowledge, the only prior research in this direction is by G\"urlebeck~\cite{Gurlebeck2012,Gurlebeck2015}, who demonstrated that the Weyl multipoles for a static and axisymmetric configuration can be expressed as a sum of surface integrals surrounding both the matter and black hole regions. In their interpretation, since the contribution to the multipolar structure from the black hole—i.e., the ``hair''—is captured by a single parameter, they conclude that this constitutes a version of the no-hair theorem for black holes in external gravitational fields. However, we argue that this statement can be expressed in a manner closer to more traditional uniqueness results. Our results demonstrate that in a static and axisymmetric scenario, there exists at most a one-parameter family of horizon shapes that can coexist with an external gravitational field (or a given matter configuration). This parameter represents an overall size for the black hole (i.e. it represents its mass). For the environment to be consistent we have shown that it is necessary 
that it exhibits a gravitational equilibrium condition, as worked out in detail in Section~\ref{Sec:NoHair}. 
Therefore, we can say that from a pure geometric (or kinematic) point of view one can have any shape for the black hole horizon. All the uniqueness results come from the dynamical requirements imposed by the vacuum Einstein equations, along with the boundary conditions of the entire gravitating system, one of which is asymptotic flatness being one of them. 

At this point, we find it necessary to clarify a common misconception about the deformability of black holes, leveraging the results of G\"urlebeck. Love numbers, well established in Newtonian gravity and with some working definitions in general relativistic contexts~\cite{Binnington2009,Poisson:2020vap}, are known to vanish for static black holes~\cite{Damour2009,Damour2009b,Kol2011,Charalambous:2021mea}. For a static object with a specific multipolar structure deformed by an external gravitational field, Love numbers quantify the extent to which the object's contribution to the multipolar structure deviates in response to the strength of the external field. As we have shown, this interpretation holds, at least in the static and axisymmetric case under consideration here. However, the vanishing of Love numbers must not be confused with a non-deformability of the cross-sectional geometry of event horizons. Indeed, as the solutions presented here demonstrate, it is evident that the intrinsic geometry of the horizon cross section is deformed due to the presence of external matter. Furthermore, this suggests that one might be able to define some different non-vanishing ``Love numbers" to specify how the contribution of the external matter affects the multipolar shape of the deformed horizons. 

Our explanation of why black holes cannot be distorted from the inside is somewhat technical, so here we offer a more intuitive, physical interpretation. A lump of matter can resist its own gravitational pull by generating internal positive pressures that counteract it. When an object is not very compact, this internal pressure largely determines its shape. For example, objects such as asteroids with very low compactness can take on very irregular, non-spherical shapes. However, as an object’s compactness increases, gravity becomes a dominant factor in shaping it, and non-rotating objects tend to become more spherical. This spherical shape minimizes the object’s gravitational self-energy. When the compactness of an object surpasses the black-hole threshold (i.e., when an event horizon forms), any structure capable of resisting gravitational collapse would require violating energy conditions—either by creating negative pressures or negative energy densities. For instance, one could construct a spherically symmetric black holes that only differ from the Schwarzschild black hole in an inner core located in the black hole region~\cite{Frolov1989}. In fact, any non-spherically symmetric regular black hole would require external matter to sustain the deviation from spherical symmetry. This can be understood as follows: consider a geometry with a non-vanishing $T_{\mu \nu}$ in an inner core that is not spherically symmetric but depends continuously on a parameter representing its compactness. As the configuration becomes progressively more compact, there will be a point where a horizon forms. However, our analysis indicates that such a horizon would inevitably leave some matter outside. If the horizon were to enclose all the matter content, it would necessarily result in a spherically symmetric horizon.

Finally, in this paper we have explored how the no-hair results are modified when relaxing the assumption of the presence of a regular horizon, replacing it with a minimum redshift surface at which the Kretschmann scalar is demanded to be smaller than Planckian. In the static and axisymmetric case considered here, we have been able to constrain all the possible ultracompact structures, including deformations of spherically symmetric ultracompact stars by requiring that the Kretschmann scalar does not reach trans-Planckian values. For deviations from the Schwarzschild metric described by an inhomogeneity function in the direction of the symmetry axis, we find that these inhomogeneities (which are equivalent to non-sphericities) have to die off as the structure becomes more and more compact. In the domains of validity where our analyses intersect with those in~\cite{Barcelo2019} and~\cite{Raposo2019}, we find essentially the same results: that deviations from sphericity have to die off with powers of the compactness parameter. In addition, our analysis has been able to disentangle the role played by the Curzon family of metrics, unveiling why some compactness limits do not lead to the Schwarzschild geometry. We believe that this work would help to better understand the physics behind uniqueness theorems and to better appreciate the additional difficulties in the general static and stationary cases.

\acknowledgments{
GGM thanks Béatrice Bonga, Alejandro Jim\'enez Cano, Ariadna Ribes Metidieri, Miguel S\'anchez, Stefano Liberati, Jos\'e M.M. Senovilla and Ra\"ul Vera for very enlightening discussions and Robert Bryant for helpful correspondence. Financial support was provided by the Spanish Government through the Grants No. PID2020-118159GB-C43, PID2020-118159GB-C44, PID2023-149018NB-C43 and PID2023-149018NB-C44 (funded by MCIN/AEI/10.13039/501100011033), and by the Junta de Andaluc\'{\i}a through the project FQM219. GGM is funded by the Spanish Government fellowship FPU20/01684. CB and GGM acknowledge financial support from the Severo Ochoa grant CEX2021-001131-S funded by MCIN/AEI/ 10.13039/501100011033. RCR acknowledges financial support through a research grant (29405) from VILLUM fonden. LJG also acknowledges the support of the Natural Sciences and Engineering Research Council of Canada (NSERC).}

\appendix 

\section{Topology of event horizons}
\label{App:Topology_Horizons}

\begin{figure} 
\begin{center}
\includegraphics[width=0.35 \textwidth]{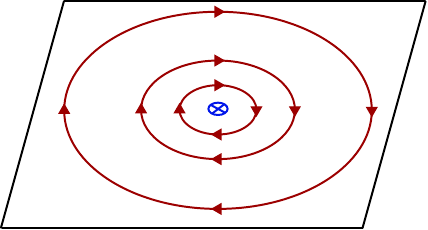}
\caption{The index of the axial Killing vector field $\bm m$ is always $+1$ since it generates always orbits that can be mapped to a circle $S^1$, generating a rigid rotation around the zero of the vector, which is fixed by the action of the isometry group.}
\label{Fig:IndexVector}
\end{center}
\end{figure}

Every cross section  $S$ of the horizons considered here is a compact, orientable (since it is a boundary) surface with a positive definite metric, and an axial Killing vector field $\bm m$. According to the Poincar\'e-Hopf theorem~\cite{Milnor1997} the sum of the indices of the zeros of $\bm m$ should be equal to the Euler number of the cross section:
\begin{align}
    \sum_{i} \text{index}_i (\bm m) = \chi (S) = 2  - 2g(S),
\end{align}
with $g(S)$ being the genus of the surface $S$. The vector field $\bm m$ can have at most isolated zeros. Given that $\bm m$ is the Killing vector generating axial symmetry, we have that the index of those zeros is always $+1$, as they need to behave locally around the zero as rotations since the flow of $\bm m$ leaves the zero as a fixed point (Fig. \ref{Fig:IndexVector})
Hence, $2 - 2g \geq 0$ and $g \leq 1$, so it can only take the values $g = 0$, corresponding to a spherical horizon where the Killing vector has two zeros at the two poles of the sphere; and $g = 1$, corresponding to a toroidal horizon corresponding to the vector having no zeros. 
\begin{figure}[H]
\begin{center}
\includegraphics[width=0.35 \textwidth]{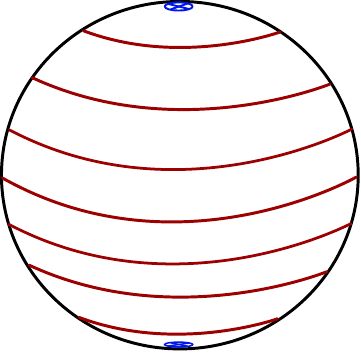}
\includegraphics[width=0.35 \textwidth]{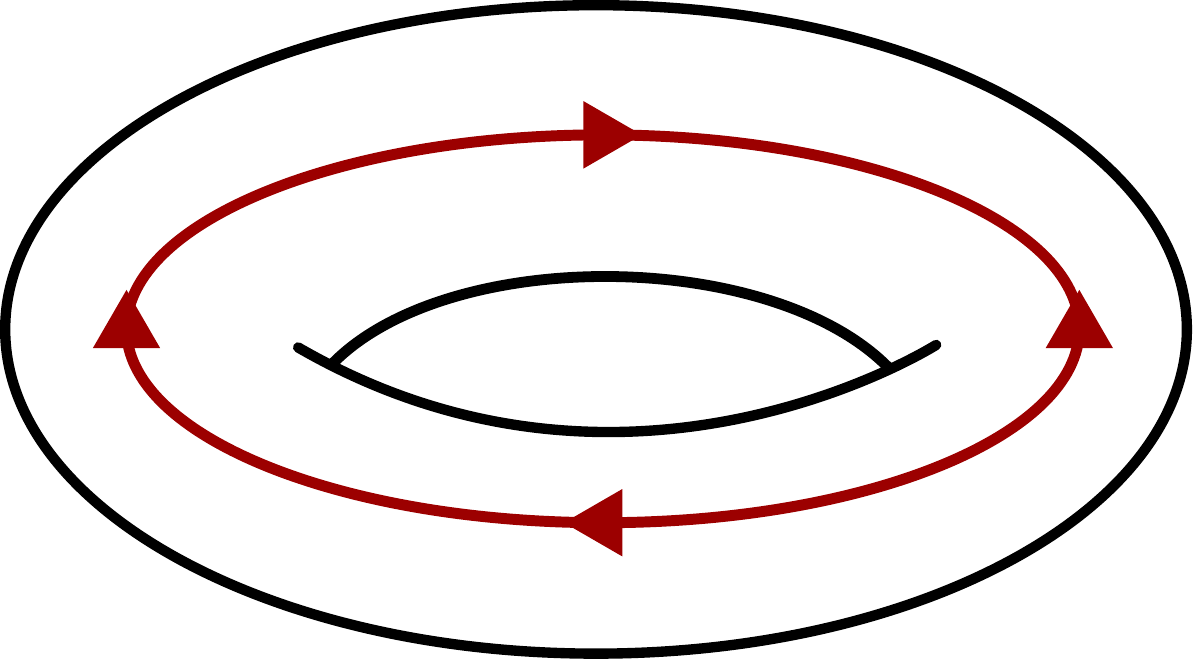}
\caption{Here we depict the two posible configurations for the cross sections of the horizon: a two-sphere with the two poles at which the vector vanishes and the toroidal configuration.}
\label{Fig:Topologies}
\end{center}
\end{figure} 
One may wonder under which conditions a toroidal cross-section is possible, specifically whether there exists a type of matter content that can distort the black hole horizon to such an extent that its topology is that of a torus rather than a sphere. First, we have to take into account the fact that we need violations of the energy conditions  in the external region according to the theorem by Hawking (see Prop. 9.3.2 of ~\cite{HawkingEllis1973}), or the one by Galloway in which some global assumptions are relaxed~\cite{Galloway1993}. See also~\cite{Chrusciel1994,Jacobson1994} for further developments.

The construction of Geroch-Hartle can be applied, but now taking the following metric as a reference metric
\begin{align}
    ds^2 = - r^2 /4m^2 dt^2 + 4 dr^2 + 4 dz^2 + 4 m^2 d \varphi ^2,
\end{align}
with the following identifications on the Riemannian three-space described by $(r,z,\varphi)$: $(r,z=m,\varphi) \sim (r,z = -m, \varphi + \alpha)$, with $m$ and $\alpha$ any real numbers, and $(r,z,\varphi) \sim (r,z,\varphi + 2 \pi)$. This geometry is flat, as can be checked by directly computing the Riemann tensor (actually, it is the flat spacetime metric in Rindler coordinates). It corresponds to the solutions $U_T = \log (r /2m)$ for the ALP, which implies that $V_T = \log (r/m)$ by Eqs.~\eqref{Eq:DrV}-\eqref{Eq:DzV}. 

In order for this metric to be continued to an asymptotically flat region we need   to cut this metric at a finite $r = R$, constituting an outer boundary (with the topology of a torus) of what we can call the inner region  $I$. Then, we can choose an asymptotic region that has a spherical inner boundary $S^2$; we call this the external region $E$. The whole geometry is obtained by matching this inner region $I$ and the external region, through a three dimensional Riemannian compact manifold with a disconnected boundary composed of the torus and the sphere. This spacetime needs to violate energy conditions somewhere. 

\begin{figure}
\begin{center}
\includegraphics[width=0.32 \textwidth]{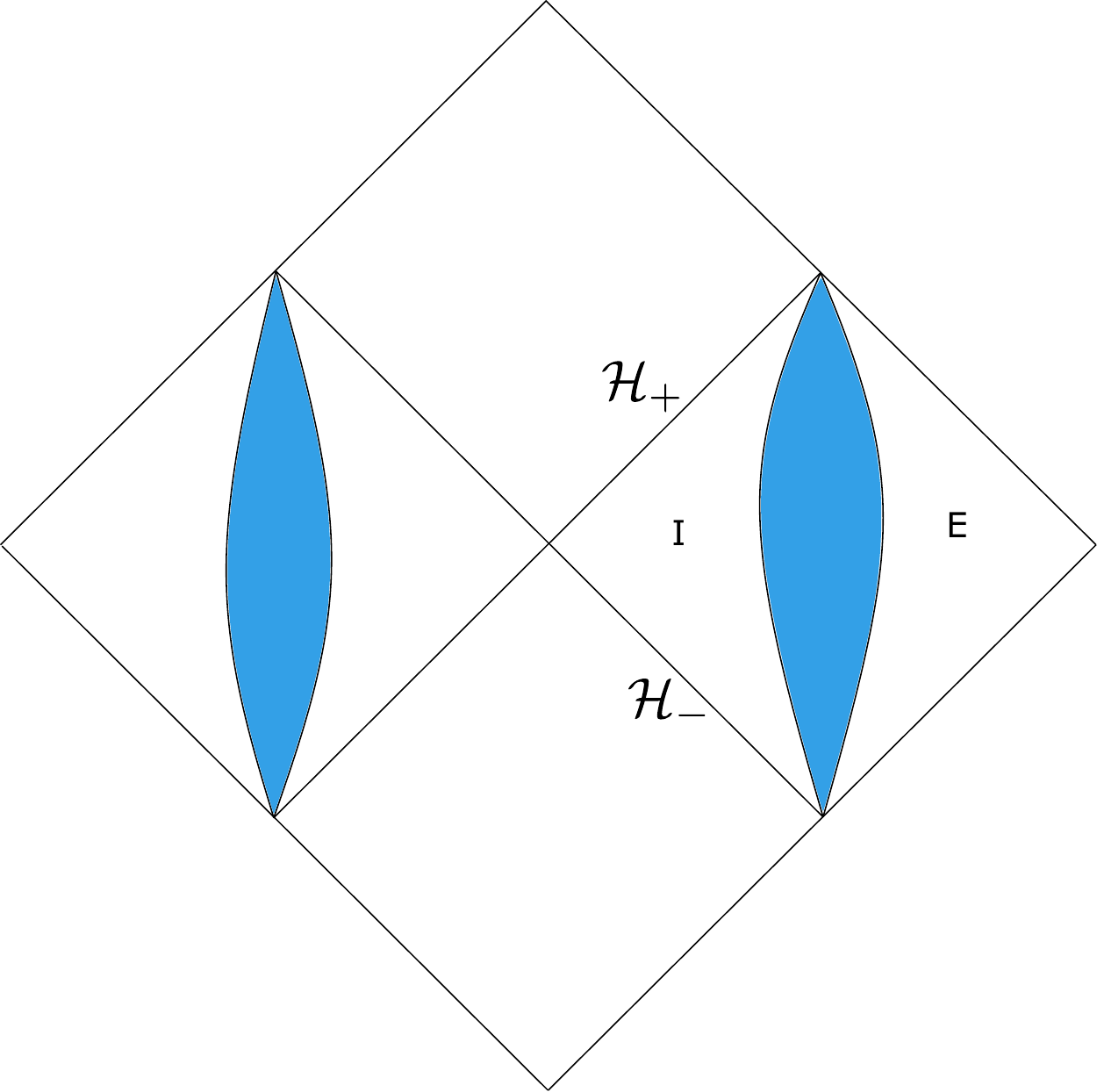}
\caption{Causal structure of the toroidal black hole. $\mathcal{H^{+}}$ and $\mathcal{H^{-}}$ represent the future and past horizons, and $I$ and $E$ the internal and external regions of the metric we began with before extending through the horizon. The time reversal of this part is included in the left diamond of the diagram. Furthermore, the blue shaded regions correspond to the region interpolating between the inner region (where every point in the diagram represents a torus $S^1 \times S^1$), and the external region, where every point represents a two-sphere $S^2$. Energy condition violations need to occur somewhere in the blue shaded region or the external region $E$.}
\label{Fig:Causal_Structure_Torus}
\end{center}
\end{figure}

We can now consider the analytical continuation to the region inside the black hole. Actually, we can perform a change of coordinates to consider the maximal analytic extension of this metric. We can make the change of coordinates 
\begin{align}
    & R^2 - T^2 = 4 r^2, 
    & T/R = \tanh [t/(8m)],
\end{align}
so that the metric is explicitly flat
\begin{align}
    ds^2 = - dT^2 + dR^2 + 4 dz^2 + 4m^2 d \varphi^2. 
\end{align}
In this way, we can extend the metric through the horizon located at $r = 0$. Furthermore, the black hole region has no singularity, the geodesics are complete since we are in flat spacetime with identifications. The structure of this black hole is similar to the standard Kruskal spacetime with the two spheres replaced by a torii and it is depicted in Fig.~\ref{Fig:Causal_Structure_Torus}. 

Similar results could be obtained for any distorted black hole, which tantamount to replacing $U_T,V_T$ with $U_T + \hat{U}, V_T + \hat{V}$, and it is expected to find the same kind of causal structure presented here~\cite{Geroch1982}. In the main text, we have restricted our analysis to spherical black holes, as toroidal ones are less well understood and likely less relevant in astrophysical contexts, given that they require violations of energy conditions in the black hole’s exterior region. Furthermore, there are not any explicit examples of known toroidal black holes with regular asymptotic regions (for some examples with singularities in the external regions see~\cite{Thorne1975,Peters1979}). 

\section{Bach-Weyl rings}
\label{App:Rings}

We can consider the solution to the Poisson equation associated with a density profile of the form 
\begin{align}
    \rho(\boldsymbol{x}) = \frac{M_{\text{ring}}}{2 \pi}  \delta (z) \delta( r - r_{\text{ring}}), 
\end{align}
which represents a ring of radius $r_{\text{ring}}$ located at the $z = 0$ plane and constant density. The solution can be expressed in terms of the Green function explicitly as 
\begin{align}
    U( \boldsymbol{x}) = -  \int d^3\boldsymbol{x'} \frac{\rho ( \boldsymbol{x'})}{\abs{\boldsymbol{x} - \boldsymbol{x'}}}.
    \label{Eq:FundamSol}
\end{align}
It is possible to analytically perform this integral and express it in terms of the elliptic function $K$~\cite{Abramowitz1964} as
\begin{align}
    U_{\text{ring}}(r,z) = - \frac{2 M_{\text{ring}} }{\pi \sqrt{z^2 + (r_{\text{ring}}+r)^2 }} K \left(  \frac{4r_{\text{ring}}r}{z^2 + (r_{\text{ring}}+r)^2} \right). 
    \label{Eq:RingPotential}
\end{align}
Everywhere except for the ring where matter is located, this function is a harmonic function, and hence it is solution to Laplace equation. We can take this as a solution to generate a solution of Eqs.~\eqref{Eq:DrV}-\eqref{Eq:DzV} of GR. This is the so-called Bach-Weyl ring~\cite{Bach2012}. Plugging $U$ into the equations for $V$, it turns out that $V$ can be found analytically~\cite{Bach2012}:
\begin{align}
    & V_{\text{ring}}(r,z) =  \frac{ M_{\text{ring}}^2  }{4 \pi^2 r r_{\text{ring}}} \left( \frac{4 r r_{\text{ring}}}{z^2 + (r_{\text{ring}}+r)^2} \right)^2 \Bigg[ - \left[ K \left( \frac{4rr_{\text{ring}}}{z^2 + (r+r_{\text{ring}})^2} \right) \right]^2 \nonumber \\
    & + 4 \frac{z^2 + (r-r_{\text{ring}})^2}{z^2 + (r+r_{\text{ring}})^2}  K \left( \frac{4rr_{\text{ring}}}{z^2  + (r+r_{\text{ring}})^2} \right)  K' \left( \frac{4rr_{\text{ring}}}{z^2 + (r+r_{\text{ring}})^2}  \right)  \nonumber \\
    & + 4 \frac{4rr_{\text{ring}}}{z^2 + (r+r_{\text{ring}})^2} \frac{z^2 + (r-r_{\text{ring}})^2}{z^2 + (r+r_{\text{ring}})^2}  \left[ K' \left( \frac{4rr_{\text{ring}}}{z^2 + (r+r_{\text{ring}})^2}  \right)  \right]^2 \Bigg] \nonumber \\
    & +  \frac{ M_{\text{ring}}^2 }{4 \pi^2 r_{\text{ring}}^2}  \left( \frac{4rr_{\text{ring}}}{z^2 + (r+r_{\text{ring}})^2} \right)^2 \Bigg[ - \left[ K \left( \frac{4rr_{\text{ring}}}{z^2 + (r+r_{\text{ring}})^2} \right) \right]^2 \nonumber \\
    & + 4 \frac{z^2 + (r-r_{\text{ring}})^2}{z^2 + (r+r_{\text{ring}})^2} K \left( \frac{4rr_{\text{ring}}}{z^2 + (r+r_{\text{ring}})^2} \right)  K' \left( \frac{4rr_{\text{ring}}}{z^2 + (r+r_{\text{ring}})^2} \right) \nonumber \\
    & - 8 \frac{z^2 + (r-r_{\text{ring}})^2}{z^2 + (r+r_{\text{ring}})^2} \frac{z^2 + r^2 + r_{\text{ring}}^2}{z^2 + (r+r_{\text{ring}})^2} \left[ K'  \left( \frac{4rr_{\text{ring}}}{z^2 + (r+r_{\text{ring}})^2} \right) \right]^2 \Bigg].
\end{align}
This function $V_{\text{ring}}(r,z)$ obeys all the requirements of regularity of the metric: it   approaches zero quadratically in $r$ as we approach the $z$-axis, while it goes smoothly to zero at spatial infinity. This solution can be simply translated in the $z$ axis by making the shift $z \rightarrow z - z_0$ above. 

We can now combine this with the Schwarzschild solution in order to find a new solution representing a black hole distorted by the gravitational field of a ring of matter, namely:
\begin{align}
    U_{\text{comb}}(r,z;z_0) = U_S(r,z) + U_{\text{ring}}(r,z-z_0).
\end{align}
We have plotted the corresponding derivative of $V_{\text{comb}}$, namely 
\begin{align}
    \partial_r V_{\text{comb}} = r \left[  \left( \partial_r  U_{\text{comb}} \right)^2 - \left( \partial_z  U_{\text{comb}} \right)^2  \right], 
\end{align}
in Fig.~\ref{Fig:Saturns_Integrands} for the ring lying on the same plane as the black hole $z_0 =0 $ and a value $z_0 \neq 0$. Notice that other solutions can be found that are generated by an infinitesimally thin ring, see for instance the ``elliptic metric'' found in~\cite{Batic2023} through the Hankel integral transform method.

\section{Curzon metric and higher multipoles}
\label{App:Curzon}

\subsection{Curzon metric}

Consider the Curzon metric~\cite{Curzon1925}, which can be expressed in terms of the Weyl line-element~\eqref{Eq:Metric}
where the functions $U(r,z)$ and $V(r,z)$ are given by
\begin{align}
    & U(r,z) = - \frac{M}{\sqrt{r^2+z^2}}, \\
    & V(r,z) = - \frac{M^2 r^2}{2(r^2 + z^2)^2}, 
\end{align}
with $M$ a real parameter. We need to distinguish the situation in which $m$ is positive and the situation in which $m$ is negative. It is convenient to change coordinates to spherical coordinates
\begin{align}
    & r = R \sin \theta, \\
    & z = R \cos \theta,
\end{align}
in terms of which the metric reads:
\begin{align}
    ds^2 = - e^{2 U} dt^2 +e^{-2U} \left[ e^{2V} \left( dR^2 + R^2 d \theta ^2 \right) + R^2 \sin^2 \theta d \varphi^2 \right],
\label{Eq:Line-Element- }
\end{align}
with 
\begin{align}
    & U(R) = - \frac{M}{R}, \\
    & V(R,\theta) = - \frac{M^2 \sin^2 \theta }{2 R^2}.  
\end{align}
The area of the constant $R$ spacelike surfaces (for a fixed time $t$) is given by the expression: 
\begin{align}
    A(R) = \int d \Omega \sqrt{g^{(2)}}
     = R^2 e^{2M/R} \int_0^{\pi} d \theta \int^{2 \pi}_0 d \varphi \sin \theta e^{-M^2 \sin^2 \theta/ (2R^2)},
\end{align}
which upon performing the integral over $\varphi$ and   rearranging  terms reads
\begin{align}
    A(R) = 2 \pi R^2 e^{2M/R - M^2/(2 R^2)} \int_{-1}^1 d (\cos \theta) e^{M^2 \cos^2 \theta / (2R^2)}.
\end{align}

We can absorb $M$ into the radial coordinate by performing the change $R \rightarrow  |M| R$, since it only corresponds to a choice of units. Also by parity arguments we can multiply by two and integrate between $0$ and $1$ to obtain
\begin{align}
    & A(R)/M^2 = 4 \pi R^2 e^{2/R - 1/(2R^2)}  \int_{0}^1 d u e^{ u^2 / (2R^2)} \quad M >0, \\
    & A(R)/M^2 = 4 \pi R^2 e^{-2/R - 1/(2R^2)}  \int_{0}^1 d u e^{ u^2 / (2R^2)} \quad M<0.
\end{align}
We will also use the Kretschmann scalar of the metric
\begin{align}
    \mathcal{K} = -\frac{8 M^2 e^{\frac{2 M \left(M \sin ^2(\theta )-2 R\right)}{R^2}} \left(3 R^2 \left(M^2 (\cos (2 \theta )-3)+4 M R-2 R^2\right)-2 M^3 \sin ^2(\theta ) (M-3 R)\right)}{R^{10}}.
\end{align}

\paragraph*{\textbf{Positive M.}}
The limit $R \rightarrow 0$ is highly directional for the curvature. As we approach $R=0$ through a direction of fixed $\theta=\theta_0$, we have that the  
\begin{align}
    & \lim_{R \rightarrow 0} \mathcal{K}(R,\theta_0) = \infty, \quad \forall \theta_0 \neq 0, \pi, \\
    & \lim_{R \rightarrow 0} \mathcal{K} (R,0) = \lim_{R \rightarrow 0} \mathcal{K} (R,\pi) = 0.
\end{align}
It is quite surprising that as we approach the (at least coordinate) singularity located at $z = r = 0$, the Kretschmann scalar may vanish or not, depending on the direction through which we approach it. This is an archetypal example of directional singularity. We have depicted, see Fig.~\ref{Fig:Curvatures_Redshift}, the equipotential surfaces and the surfaces of constant Kretschmann scalar for the sake of comparison with their Schwarzschild counterpart. 

\begin{figure}[H]
\begin{center}
\includegraphics[width=0.75 \textwidth]{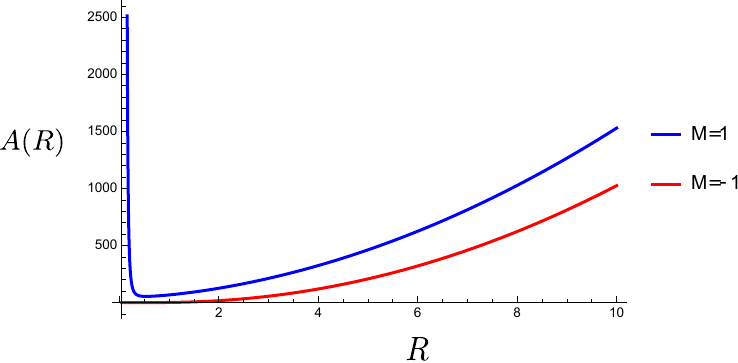}
\caption{We are depicting here the area function as a function of $R$ for $M=1$ in blue and $M=-1$ in red. Although for $M=1$ the area decreases as we decrease $R$ until we reach a minimum $R \simeq 0.538905$ and it increases as we decrease the value of $R$ from that value on. This contrasts with the area for $M = -1$ which is a monotonically decreasing function.}
\label{Fig:AreaFunction}
\end{center}
\end{figure}

Furthermore, we can take a look at the area function to see that the constant $R$ surfaces decrease in area as we decrease the value of the coordinate $R$ until we reach the surface at which $R \simeq 0.538905M$, see Fig~\ref{Fig:AreaFunction}. From that point on, we have that the area of the surfaces increases without a bound. In the limit $R \rightarrow 0$, the area goes to infinity~\cite{Stachel1968}.
All the discussion until now has been about the nature of the constant $t$ slices of the metric, i.e., an analysis of the spacetime from the point of view of approaching the singularity through spacelike directions. However, from the point of view of GR, null and timelike geodesics are the most important ones since they are relevant from the point of view of the physical observers.

Regarding the surface gravity, we  it can be obtained as
\begin{align}
    \kappa^2 = - \frac{1}{2} \lim (\nabla_A \chi_B) (\nabla^A \chi^B),
\end{align}
where the limit represents that we are evaluating the object on the horizon. For the line element in Eq.~\eqref{Eq:Metric}, it is shown by Geroch and Hartle, see Eq.(3.4) from~\cite{Geroch1982}, that the surface gravity admits the following expression as the horizon is approached:
\begin{align}
    \kappa^2(r,z) = \lim e^{4 U (r,z) - 2 V(r,z)} \left[ \left(\partial_r U(r,z) \right)^2 + \left( \partial_z U(r,z) \right)^2 \right]. 
\end{align}
If we plug in the functions for the Curzon metric, we find that the leading order behavior at $r=z=0$ as we we approach it through an arbitrary direction $r = R \sin \theta$ and $z = R \cos \theta$ is given by
\begin{align}
    \kappa^2 (\theta) = \frac{M^2 }{R^4}\exp{\frac{M \big(M [-\cos (2 \theta ])+M-8 R\big)}{2 R^2}} + \text{subdominant terms}.  
\end{align}
Taking the square root and taking into account that the term in the exponential is such that as long as $\cos \theta \neq 1$ the argument is positive, we find that:
\begin{align}
    \kappa (\theta) \rightarrow + \infty \qquad \theta \neq 0, \pi.
\end{align}
If we approach the point through the $r = 0$ direction, namely along the $z$-axis which we recall is the direction along which the Kretschmann remains finite, we find that the surface gravity is given by:
\begin{align}
    \kappa_{\text{z-axis}} = \lim _{z \rightarrow 0} \frac{M e^{- {2 M}/{\left| z\right| }}}{z^2} = 0,
\end{align}
as it is to be expected, given that the Kretschmann scalar remains finite, and there is a piece in the Kretschmann scalar that contains derivatives of $\kappa$ along the directions transverse to the horizon. 
\paragraph*{\textbf{Negative M.}}

This case is much simpler and it qualitatively represents the same situation that we have in the Schwarzschild metric with negative mass parameter. Namely, the $R=0$ ``point" represents a naked timelike singularity that cannot be hit in finite proper time and the area of the surfaces in this case vanishes as $R \rightarrow 0$. Let us see this in detail.

\begin{figure}[H]
\begin{center}
\includegraphics[width=0.45 \textwidth]{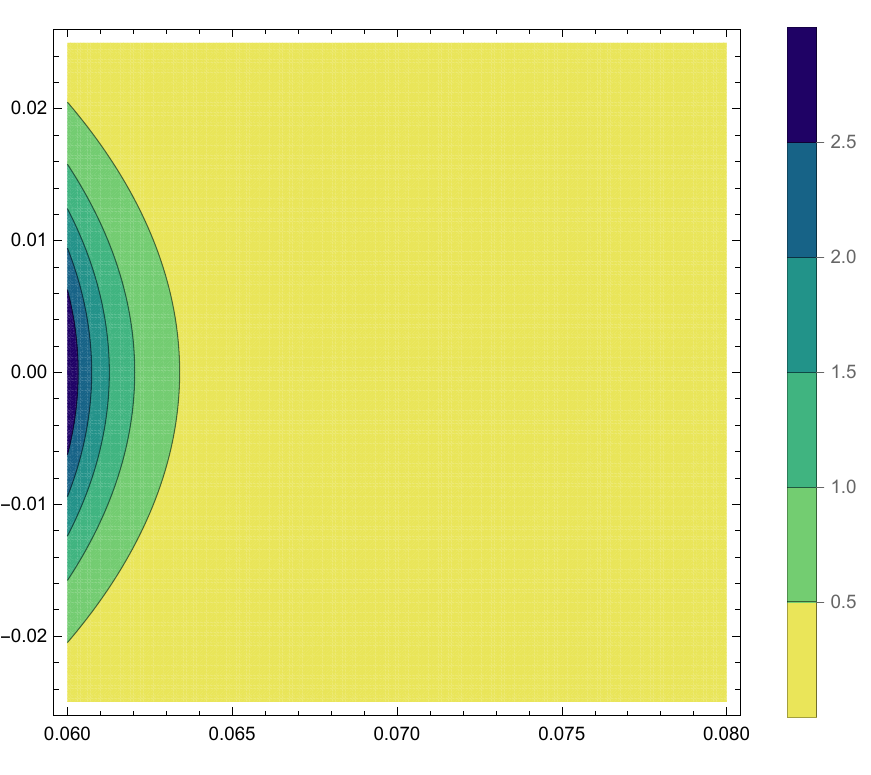}
\includegraphics[width=0.45 \textwidth]{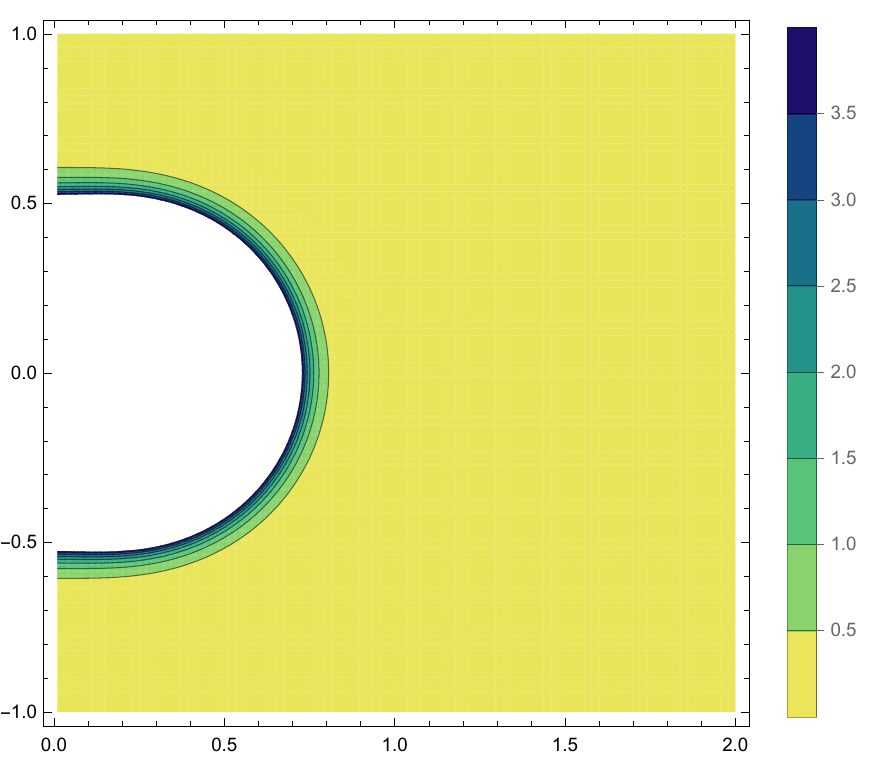}
\caption{We depict here in the left panel the constant redshift surface in the $(r,z)$ coordinate for the Curzon solution with negative mass and the Kretscmann scalar on the right panel. We normalize the graphics so that the maximum is of $\order{1}$.}
\label{Fig:Curvatures_Redshift_Negative}
\end{center}
\end{figure}

First of all, it is easy to see that  the limit
\begin{align}
    \lim_{R \rightarrow 0} \mathcal{K} (R, \theta_0),
\end{align}
is unbounded independently of the direction along which we take it. In the case of positive mass, the prefactor $e^{2M(M \sin^2 \theta_0 - 2R)/R^2}$ was divergent along every direction, except those directions $\theta_0 = 0, \pi$ for which the first term in the argument of exponential vanish and hence, we have a factor $e^{-4M/R}$ which dominates over the $1/R^{10}$ factor and makes the Kretschman approach the value zero. However, in this case, the negative mass makes the exponential to go to infinity independently of the value of $\theta_0$ and hence the Kretschmann diverges. The redshift function also blows up as we approach $R \rightarrow 0$, as it can be seen in Fig.~\ref{Fig:Curvatures_Redshift_Negative}. 

The singularity is such that it cannot be hit by any timelike geodesic. Furthermore, it cannot be hit by any reasonable observer. We follow here the discussion in~\cite{Chakrabarti1983}. An observer will be described by a trajectory that is a timelike curve $\gamma$, with $\xi^A$ its unit tangent vector pointing toward the future, which is not necessarily a geodesic, and hence it does not need to have a vanishing acceleration $a^B = \xi^C\nabla_C \xi^B$. A reasonable observer means that the integrated acceleration along the trajectory is not infinite, namely: 
\begin{align}
    \int_{\gamma} a < \infty,
\end{align}
with $a = \sqrt{a^B a^C g_{BC}}$. This condition is equivalent to the statement that there exists a function $m$ bounded from below such that 
\begin{align}
    J^C = - \xi^B \nabla_B \left( m \xi^C \right),
\end{align}
is always timelike or null (pointing toward the future). We can show this expanding its expression
\begin{align}
    J^C = - \xi^C \xi^B \nabla_B m - m \xi^C \nabla_C \xi^B = - \xi^B \nabla_B m - ma^C. 
\end{align}
We observe that $a^B \xi_B  = 0$, as it can straightforwardly be checked by taking the derivative $\xi^C \nabla_C$ in $\xi^B \xi_B = -1$, and hence if we compute 
\begin{align}
    g_{AB} J^A J^B = \left( \xi^C \nabla_C m \right)^2 - m^2 a^2,
\end{align}
asking for $g_{AB} J^A J^B \leq 0$, we get that
\begin{align}
    a \leq - \xi^B \nabla_B \log m. 
\end{align}
An integration of this, demonstrates the statement about the equivalence between the existence of $m$ and the finite acceleration. The spacetime that we are considering is static, hence we have the timelike Killing vector field with components $t^A$, hence we have $\nabla_{(A} t_{B)}=0$. As a consequence of this, we have the conserved quantity along geodesics that we can identify with the effective energy:
\begin{align}
    E = -\xi_B t^B. 
\end{align}
In principle, $\gamma$ is not a geodesic, and hence $E$ is not conserved. However, we can prove the identity 
\begin{align}
    | \xi^B \nabla_B E | \leq a E.
\end{align}
Expanding the derivative of $E$ we have $\xi^B \nabla_B E = -a_B t^B - \xi^B \xi^B \nabla_{B} t_{C}$. The last term vanishes since it is contracted with a symmetric tensor and it vanishes through the Killing equation. We have then:
\begin{align}
    -a_B t^B = a_B t_C h^{BC}, 
\end{align}
where we have introduced the projector onto the space orthogonal to the trajectory $h^{AB} = g^{AB} + \xi^A \xi^B$, and use the fact that the acceleration is always orthogonal to the vector $\xi^B$. Now we can apply the Cauchy-Schwarz inequality to find
\begin{align}
    a_B t_C h^{BC} \leq \left( a_B a_C h^{BC} \right)^{1/2} \left( t_B t_C h^{BC }\right)^{1/2} = a \left( t^B t_B + E^2 \right) \leq a E,
\end{align}
since in our conventions the timelike character of $t^B$ means that $t^B t_B \leq 0$. This leads to $\abs{\xi^B\nabla_B E} \leq aE$. If we integrate this inequality now along the curve, we conclude that $E$ needs to be finite along the curve also. We can use the triangular inequality applied to the definition of the energy leads to
\begin{align}
    E = - \xi^B t_B \geq \left( - t^B t_B \right)^{1/2} \left( - \xi^B \xi_B \right)^{1/2} = \left( - t^B t_B \right)^{1/2}. 
\end{align}
Applying this to our spacetime we have that $E$, which needs to be bounded (and hence there exists an upper bound $\Tilde{E} \in \mathbb{R}$ for it), obeys:
\begin{align}
   \Tilde{E} \geq  E \geq  (- t^B t_B)^{1/2}  = e^{ {\abs{m}}/{R}}, 
\end{align}
and hence we conclude that it is not possible to reach arbitrarily small values of the $R$ coordinate. Hence, no ``decent'' trajectory can reach the naked singularity located at $R = 0$. 

Finally, the area in this case is a monotone increasing function of $R$, approaching $0$ as we approach $R=0$, as it can be seen in Fig.~\ref{Fig:AreaFunction}.
\subsection{Higher-multipole configurations}
We can consider the solution generated by an arbitrary multipole located at the origin:
\begin{align*}
    U^{(\ell)} = - \frac{M^{(\ell)}}{(r^2 + z^2)^{\frac{\ell+1}{2}}} P_{\ell} \left( \frac{z}{\sqrt{r^2+z^2}} \right),
\end{align*}
and we can now perform the integration of the equations for $V$ in order to find the following:
\begin{align}
    V^{(\ell)} = - \frac{(\ell+1) (M^{(\ell)})^2}{2 (r^2+z^2)^{\ell+2}} \left[ P_{\ell}^{2} \left( \frac{z}{\sqrt{z^2 + r^2}} \right) - P_{\ell+1}^{2} \left( \frac{z}{\sqrt{z^2 + r^2}} \right) \right]. 
\end{align}
Actually, it is possible to perform the integral even for an arbitrary superposition of multipolar configurations:
\begin{align*}
    U(r,z) = - \sum_{\ell=0}^{\infty} \frac{M^{(\ell)}}{(r^2 + z^2)^{\frac{\ell+1}{2}}} P_{\ell} \left( \frac{z}{\sqrt{r^2+z^2}}\right),
\end{align*}
with the function $V$ reading:
\begin{align}
    V(r,z) = & - \sum_{\ell, m =0}^{\infty} \frac{M^{(\ell)} M^{(m)} (\ell + 1) (m + 1)}{(\ell + m + 2) (r^2 + z^2)^{\ell + m + 2/2}} \\ 
    \times & \left[ P_{\ell} \left( \frac{z}{\sqrt{z^2 + r^2}} \right) P_{m} \left( \frac{z}{\sqrt{z^2 + r^2}} \right) - P_{\ell+1} \left( \frac{z}{\sqrt{z^2 + r^2}} \right) P_{m+1} \left( \frac{z}{\sqrt{z^2 + r^2}} \right) \right]. 
\end{align}
We can compute the Kretschmann scalar of these functions for different multipolar configurations and all of them show a behavior that is qualitatively similar to the Curzon metric: there exist directions for approaching the $ r = z = 0$ point giving rise to a curvature singularity, and others that do not.

\bibliography{israel_biblio}

\end{document}